%% file: usenix2019_v3.1.tex
\documentclass[conference]{IEEEtran}
\IEEEoverridecommandlockouts

\usepackage[shortlabels]{enumitem}
\usepackage{filecontents}

\usepackage{graphicx}
\usepackage{xcolor}
\usepackage{colortbl}
\usepackage{tikz,pgfplots}
\usepackage{amsmath}
\usepackage{amssymb}
\usepackage{url}
\usepackage{subcaption}
\usepackage{pgfplots}
\usepackage{algorithm}
\usepackage{longtable}
\usepackage[noend]{algpseudocode}
\usepackage{tikz,pgfplots}
\usepackage{color,soul}

\input{macros.tex}

\algnewcommand\algorithmicforeach{\textbf{for each}}
\algdef{S}[FOR]{ForEach}[1]{\algorithmicforeach\ #1\ \algorithmicdo}
\begin{document}

\date{}

\title{MaMaDroid 2.0 - The Holes of control  flow  graphs}
\author{
Harel Berger\footnote{Corresponding author.}\\
Ariel Cyber Innovation Center\\ Computer Science Department,\\ Ariel University,  Ariel, Israel\\
harel.berger@msmail.ariel.ac.il
\and 
Chen Hajaj\\
Ariel Cyber Innovation Center,\\
Data Science and Artificial Intelligence Research
Center,\\
Industrial Engineering and Management Department,\\ Ariel University,  Ariel, Israel\\
chenha@ariel.ac.il
\and
Enrico Mariconti\\
 Jill Dando Institute\\  Department of Security and Crime Science,\\ UCL,  London, United Kingdom\\
 e.mariconti@ucl.ac.uk\\
 \and
  Amit Dvir\\
 Ariel Cyber Innovation Center\\ Computer Science Department,\\ Ariel University,  Ariel, Israel\\
 amitdv@g.ariel.ac.il\\
} 
\maketitle              
\begin{abstract}
	Android malware is a continuously expanding threat to billions of mobile users around the globe. Detection systems are updated constantly to address these threats. However, a backlash takes the form of \emph{evasion attacks}, in which an adversary changes malicious samples such that those samples will be misclassified as benign. This paper fully inspects a well-known Android malware detection system, MaMaDroid, which analyzes the control flow graph of the application. Changes to the portion of benign samples in the train set and models are considered to see their effect on the classifier. The changes in the ratio between benign and malicious samples have a clear effect on each one of the models, resulting in a decrease of more than 40\% in their detection rate. Moreover, adopted ML models are implemented as well, including 5-NN, Decision Tree, and Adaboost. Exploration of the six models reveals a typical behavior in different cases, of tree-based models and distance-based models. Moreover, three novel attacks that manipulate the CFG and their detection rates are described for each one of the targeted models. The attacks decrease the detection rate of most of the models to 0\%, with regards to different ratios of benign to malicious apps. As a result, a new version of MaMaDroid is engineered. This model fuses the CFG of the app and static analysis of features of the app. This improved model is proved to be robust against evasion attacks targeting both CFG-based models and static analysis models, achieving a detection rate of more than 90\% against each one of the attacks.
	
\end{abstract}

\section{Introduction}
\label{Intro}
Malicious software, a.k.a malware, is defined as a file or program that tries to damage the normal activity of a digital device. Most malware are specifically engineered to the operating system (OS) they target, as OSs tend to vary based on their hardware and functionalities. One of the popular targets of malware is the Android OS. Android application Packages (APKs), the popular executable files of Android OS, can be found in many Android markets around the world (e.g., Google play store~\cite{GooglePlay}). Many malware apps can be found on these markets that target 
the attention of unsuspecting users to assure downloading of the malicious app. For example, the Etinu malware leverages the fear of the recent COVID-19 in Asia~\cite{covid-report}. This malware stole information from incoming SMS messages, made purchases in the victim's name, and infected more than 700K users. As a safeguard for these users and many more, a vast amount of researchers and cyber experts are looking for a satisfying solution for the correct identification of malware applications~\cite{aafer2013droidapiminer,arp2014drebin,berger2021crystal,cai2018towards,chen2016stormdroid,dini2012madam,huynh2017new,enck2009lightweight,onwuzurike2019mamadroid,shabtai2012andromaly,shabtai2009detection,shabtai2014mobile,sun2016sigpid,talha2015apk,treadwell2009heuristic,venugopal2008efficient,wang2014exploring,wu2012droidmat,xu2013permlyzer}. Several studies were conducted throughout the years to mitigate the threat of Android malware, implementing methods like using heuristics of app structure and signatures~\cite{treadwell2009heuristic,venugopal2008efficient}, permissions' analysis~\cite{enck2009lightweight,wang2014exploring,xu2013permlyzer,talha2015apk,sun2016sigpid}.

Hitherto, the most popular approaches to Android malware detection are Machine Learning  and Deep Learning~\cite{aafer2013droidapiminer, arp2014drebin, onwuzurike2019mamadroid, shabtai2009detection,berger2021crystal}. A strong ML classifier is based, among others, on the raw data that must represent the domain correctly.
Specifically, for different sub-domains of malware detection, there is a different real-world distribution of malicious and benign samples. For example, in the subdomain of URL malware detection, most Internet addresses are considered malicious~\cite{maggi2013two,rahbarinia2017exploring}.
In the paper domain, Android malware, most of the apps are considered benign~\cite{G_201_sum,lindorfer2014andradar,pendlebury2018tesseract}. A recent and popular study defined the advisable ratio as 90/10 between benign and malicious Android apps~\cite{pendlebury2018tesseract}. As a result, the challenge for the Android malware classifiers is to analyze the important characteristics of malicious apps, despite the low volume of malicious apps in the dataset. 

Even an ML classifier trained on the right amount of benign and malicious apps is not 100\% accurate on any input sample. It was proven by Goodfellow et al.~\cite{goodfellow2014explaining} that some ML classifiers are susceptible to manipulations. These manipulations are called~\textit{adversarial examples}. Adversarial examples are created when an attacker manipulates malicious samples so that the sample will be misclassified as benign and vice versa~\cite{grosse2017adversarial,kuppa2019black,yuan2019adversarial}. In turn, \textit{evasion attacks} are intelligent attacks in which the adversary manipulates malicious instances, such that they will be wrongly classified. 
Evasion attacks that change the physical malicious instance are called problem-based evasion attacks. On the other hand, evasion attacks that manipulate the extracted feature vector of an instance are called feature-based evasion attacks.

To address the threats that arise from evasion attacks on APKs, this work follows one of the well-known Android malware detection systems, MaMaDroid~\cite{onwuzurike2019mamadroid}. MaMaDroid is based on Control Flow Graph (CFG)~\cite{10.1145/390013.808479}, where a CFG is a representation of a program using graph notation. This representation depicts all paths that might be traversed through the program while executing it. 

The contribution of this paper is threefold. First, a full evaluation of the portion of instances on the training set that is benign on the classifier's detection rate is presented. Second, three innovative problem-based evasion attacks against MaMaDroid are introduced. The evasion attacks are thoroughly evaluated based on multiple metrics (e.g., robustness), and a varied set of models. Finally, 
insights from this analysis motivated the creation of a new version of MaMaDroid, MaMaDroid2.0. The new version incorporates an extended feature set. MaMaDroid2.0 was evaluated against our novel evasion attacks and other known evasion attacks. Compared to the original MamaDroid model ~\cite{onwuzurike2019mamadroid} which results in an evasion robustness of less than 30\%, the new model results in an evasion robustness of more than 90\%.
 
The remainder of this paper is organized as follows. First, related work is presented and discussed in Section~\ref{related}. MaMaDroid, the targeted system, is presented in Section~\ref{machines}, together with the attacker models. In Section~\ref{attacks}, the attacks are presented. Next, in Section~\ref{eval}, metrics and evaluations are discussed.  The new version of MaMaDroid, MaMaDroid2.0, is explored in Section~\ref{mama2}. A discussion about the evasion attacks and mitigation techniques can be found in Section~\ref{diss}, along with the conclusion of the paper.  Table~\ref{abbrev_table} is an abbreviations table. 

\section{Related Work}
\label{related}
In Section~\ref{detection systems label}, four main approaches in the field of Android malware detection systems are described. Additionally, evasion attacks leverage weak spots in ML-based Android malware detection systems. 
Evasion attacks take two forms: problem-based attacks and feature-based attacks. Problem-based attacks~\cite{alzantot2018generating,apruzzese2018evading,dang2017evading,demontis2017yes,li2018textbugger,rastogi2013droidchameleon,zheng2012adam} include modification of the samples. This is the type of attack implemented in this study. Feature-based attacks~\cite{pkdd2013,carlini2017towards,chen2018droideye,shahpasand2019adversarial,shao2019multi,xu2020towards,zikratov2017formalization} map the sample into a feature vector and modify the values of the features. Feature-based attacks are easier to implement than problem-based attacks. The feature-based attacks can be generated automatically by an ML~\cite{aydogan2015automatic,chen2017adversarial,hu2017generating,ming2015replacement,zhao2019unsupervised}. Still, the correlated code in the sample that needs to be changed according to these attacks may severely damage the functionality of the sample~\cite{chen2018android,pierazzi2020problemspace,salem2018repackman,berger2020evasion}. Therefore, feature-based evasion attacks are less realizable. This study implements problem-based evasion attacks. Therefore, a review of significant works on problem-based evasion attacks on Android malware detection systems is presented in Section ~\ref{evasion-attacks-review}.

\subsection{Android Malware Detection Systems}
\label{detection systems label}
This section surveys well-known Android malware detection systems following four main approaches. The first approach is static analysis, which gathers significant strings from the Smali code files and Manifest file. The second strategy is based on the control flow graph (CFG) of the application, which traces the order of the API calls used in the app. The behavior of the app is inspected in the third approach, which gathers information on the behavior of the Android OS during the run of the app, along with network packets sent and received, etc. The last approach analyses bytecode sequences. Several Hybrid systems are discussed as well.

Probably one of the honored Android malware detection systems using static analysis is Drebin~\cite{arp2014drebin}. Drebin gathers 8 types of static features, with a specification of two main components of the APKs. Drebin extracts permissions requests, software/hardware components, and intents from the Manifest file. From the Smali code, it extracts suspicious/restricted API calls, used permissions in the app's run and URL addresses. Sec-SVM~\cite{demontis2017yes} is an improved version of Drebin, using a more evenly weighted learning model. Other versions of Drebin include a Factorization Machine~\cite{li2019android} and a DNN~\cite{AdversarialDeepEnsemble}. The DroidAPIMiner~\cite{aafer2013droidapiminer,droidapiminber_code} is similar to Drebin, in means of static analysis and feature types. The API calls and permissions are the feature set of this detection machine. The authors of this research performed a dataflow analysis for frequently used API values and package names. The static analysis detection systems look for several features from the APK. Enumeration of these features is easy and efficient. However, since most of the static features can be easily understood, they can also be  automatically manipulated without many efforts, like in~\cite{demontis2017yes,maiorca2015stealth,meng2016mystique,rastogi2013droidchameleon}.

A more robust approach is found in MaMaDroid~\cite{onwuzurike2019mamadroid}, which extracts the Control Flow Graph (CFG) of an application as a base for its feature set. MaMaDroid generates a tree of API calls based on family and package names. Then, the detection system analyzes the API call sequence performed by an app by each mode - family or package, to model its true nature. A similar approach to detect malicious third-party use in apps was used by Backes et al.~\cite{backes2016reliable}. Function and app profiling, such as types and parameters, were used to identify third-party libraries and different versions of the same library. Zhiwu et al.~\cite{xu2018cdgdroid} used CFG along with data flow to characterize Android apps, along with a CNN model to predict the labels of new samples. An extension was suggested in~\cite{zhiwu2019android}, where the same technique was used using n-grams to identify malware families. Monitoring the sequence of functions and API calls inside the app seems like a great idea. Changing the flow of the app is more complicated, since the CFG may be complex and full of details. Furthermore, changing the order of API calls of the app may damage the malicious activity of the app. However, several evasion attacks manipulate the flow of the app and succeed in deceiving this kind of detection system, such as~\cite{chen2018android,ikram2019dadidroid,maiorca2015stealth,piao2016server,sun2014nativeguard}.

The third approach tries to map the traces of a malicious app, thus acknowledging the behavior of such apps, using several operating systems and communication features, as was introduced in Andromaly~\cite{shabtai2012andromaly}. A similar approach was taken by Shabtai et al.~\cite{shabtai2014mobile}. These detection systems focused on network usage by measuring the network
traffic patterns in a host device running an app. The authors learned the statistics of network packets a user sent and received, the RTT, etc. Another research using a similar strategy is Shabtai et al.~\cite{shabtai2010intrusion}. The authors aggregate data during an app run, including user interactions like clicks and touches and OS behavior such as the CPU and network usage. Saracino et al.~\cite{saracino2016madam} found the correlations between features at four levels: kernel, application, user and package, to identify malicious applications. Wang et al.~\cite{wang2021android} explored the number of pages on virtual machines, the change of states between tasks, etc. Behavioral detection methods are based on the nature of the Android device while running various apps. These behaviors may depend on the app and are therefore hard to generalize. Therefore, they are not a complete solution to malware detection. 

Bytecode inspection is the last approach to Android malware detection. Dalvik Bytecode Frequency Analysis~\cite{kang2013android} is one example, which looks for popular Dalvik Bytecode instructions of malware apps. Sequences of Dalvik Bytecode instructions  were also explored by TinyDroid~\cite{chen2018tinydroid}. The authors of this system gathered families of Bytecode instructions under a single symbol and used n-grams~\cite{damashek1995gauging} to create the feature set. 
Yang et al.~\cite{yang2015appspear} extracted the bytecode file from the APK and converted the Bytecode to a matrix. This matrix was analyzed by a CNN. Bytecode inspection is a heavy method and ambiguous for the human eye, as it is not a convenient programming language. 

A hybrid detection system was suggested by Martín et al.~\cite{martin2019android}. This system fused the static and dynamic analysis of Android apps. The authors combined the transitions between execution states (dynamic) and the inspection of API calls (static). A combination of static analysis of permission requests and dynamic testing of their derived API calls was introduced in~\cite{wang2015reevaluating}. MARVIN~\cite{lindorfer2015marvin} inspects the nature of Android apps through static analysis of their design, certificates, etc. In addition, a dynamic analysis of behavioral activity is processed. The combination of static analysis of permissions and intents and dynamic analysis of network traffic was introduced by Ding et al.~\cite{ding2021hybrid}. A hybrid solution seems the best option to identify malicious apps. However, each method that is added as another layer of processing consumes time and resources from the host device.

\subsection{Problem-Based Evasion Attacks}
\label{evasion-attacks-review}
This section targets the problem-based evasion attacks against malware detection systems. Problem-based evasion attacks are categorized into three forms of attacks. The first uses camouflage to conceal incriminating strings and values contained in the app, by encryption and obfuscation. Next, the second incorporates noises to the app; e.g., uncalled functions. At last, the third form tries to alter the behavior of the app. It reviews the flow of the original app and manipulates the code of several function calls.

The first course of action in evasion attacks is to conceal specific suspicious components of the app. One well-known example of a concealment effort is with the help of encryption or obfuscation. Demontis et al.~\cite{demontis2017yes} obfuscated suspicious strings, packages, and API calls. Another example of concealment is packing an app inside a fellow app. DaDidroid~\cite{ikram2019dadidroid} explored a similar approach as Demontis et al.~\cite{demontis2017yes} using packing and obfuscation.

Reflection allows a program to change its nature at runtime. It is another classic evasion technique. Rastogi et al.~\cite{rastogi2013droidchameleon} presented an attack, which mixes the Demontis et al.~\cite{demontis2017yes} obfuscation method with the addition of the reflection approach.

Another form of problem-based evasion attack includes adding noise to the app. These noises mislead the classifier's labeling process. An example of noise addition can be a stub function/code injection. A stub function is a non-operational function, that does not do anything. However, it changes the original flow of an app. An example of a stub function addition is Android HIV~\cite{chen2018android}, where the authors implemented non-invoked suspicious functions against the Drebin classifier and a stub function injection against the MaMaDroid classifier. A recent example of this kind of attack is at~\cite{pierazzi2020problemspace}, where the authors implanted benign snippets of code in malicious apps to evade the Drebin and Sec-SVM~\cite{demontis2017yes} classifiers. Rosenberg et al.~\cite{rosenberg2018generic} generated an evasion attack against Android malware detection systems using API call manipulation. Three methods were used: Addition of non-operational functions to the application, obfuscation of strings, and encoding of short API calls. Cara et al.~\cite{cara2020feasibility} added non-invoked classes to the end of functions to a.

The last approach is changing the app flow. One of the ways to implement this approach is by function outlining/inlining. In function outlining, the attacker breaks a function into smaller code snippets. In function inlining, the adversary replaces a function call with the entire function body. This technique was implemented in Droidchameleon~\cite{rastogi2013droidchameleon}, which incorporated function outlining in its evasion attacks.
Another option to break the app flow is stub function, as in~\cite{chen2018android,piao2016server,sun2014nativeguard}, resulting in an ML misclassification of a malicious app.

\section{MaMaDroid and Attacker Models}
\label{machines}
In this section, the targeted system MaMaDroid~\cite{onwuzurike2019mamadroid,onwuzurike2019mamadroidold} is presented, followed by the attacker models.

\subsection{MaMaDroid}
MaMaDroid is an Android malware detection system, introduced in 2017 by Onwuzurike et al.~\cite{onwuzurike2019mamadroid}. This detection system extracts features from the Control Flow Graph (CFG) of an APK sample. It enumerates abstracted API calls to capture the behavioral model of the app. MaMaDroid operates in two modes: family and package mode. For example, the API call \textbf{android.util.Log-$>$d()} is abstracted as \textbf{android.} in the family mode, and \textbf{android.util} in the package mode. Packages or families, which are defined by the app's developer or obfuscated are abstracted as \textit{self-defined} and \textit{obfuscated}, respectively. As the structure of the evasion attack is similar for both modes, and the family mode results in lower processing time and memory, the family mode was chosen for this analysis. The structure of the evasion attack is similar for both modes. 

MaMadroid creates the features for the learning algorithm in the following manner: First, it extracts the CFG from the app. Then, it gets the sequences of API calls. Next, the APIs are abstracted using one of the modes.
At last, it constructs a Markov chain~\cite{brooks2011handbook,geyer1992practical,marjoram2003markov}, with the
probabilities of transition between any family/package. These probabilities are used as the features. For example, \textbf{androidTojava} is the feature that resembles the probability of transition between the android family to the java family. For a full description of the MaMaDroid  classifier, see~\cite{onwuzurike2019mamadroid} (implementation is available at~\cite{mama_implementation}). 

\subsection{Attacker Models}
\label{model}
This section describes two attacker models, which depict the embedded knowledge each attack holds. The first model, named as \textbf{Feature set Access (FA)}, depicts a gray-box attacker that knows the feature set of the targeted system (as the attacker model in~\cite{spooren2019use} and attack scenario F in~\cite{chen2018android}). The second model is the \textbf{Statistics Access (SA)}, which is a white-box attacker, and can access the feature set and the training data (as attack scenario FB in~\cite{chen2018android}).

\section{Evasion Attacks}\label{attacks}
Based on the attacker models, a set of evasion attacks is engineered that transfers the embedded knowledge of the defense model to a manipulated malicious APK that will be classified as benign. The idea behind these attacks is to break and change the structure of the sample so that the detection system would not recognize the manipulated samples as malicious. 

The three evasion attacks described in this section are variants of a general attack that will be termed the Structure Break attack. However, before the description of the Structure Break (StB) general algorithm, an explanation of the \textit{mode elements} is provided. These items are vital to each one of the attacks, and specifically engineered for each one of them according to their attacker model. The \textit{mode elements} are discussed in Section~\ref{mode_elem}. Then, the Structure Break (StB) general algorithm is described in Section~\ref{mama_att_algo}. At last, the variants are described in Section~\ref{ev_attacks}.

\subsection{Mode Elements}
\label{mode_elem}

 Mode elements are a subset of the feature set of the specific mode MaMaDroid analyzes. For example, for the family mode, android., java. and xml. are a part of the feature set of the family mode and therefore can be picked for the Mode elements of the family mode. In this work, the mode elements are engineered in three ways:
 \begin{enumerate}
     \item Randomly - Randomly picking a subset of the feature set. Specifically, several families like android. java. and xml can be picked from the family mode feature set.
     \item Statistically - Using the statistics of the given data. Given data can be the train and test sets, or just the test set. The given data is analyzed, to produce statistics on the elements. If the training data is given, the elements that will be picked are the ones that hold high values in the benign data, and low values in the malicious data. For example, if the java family's features (javaTojava,javaToandroid,etc.) in the benign data are high values, and also low values in the malicious data, then java will be picked for the mode elements. The idea is to try and mimic the behavior of the benign samples. If only the test data is given, the least popular elements are chosen. The test data includes malicious samples only, as this work analyzed the effect of evasion attacks against the targeted system. For example, if the java family's features (javaTojava,javaToandroid,etc.) in the test data are in low volume, then java will be picked for the mode elements. The idea behind this pick is to try and move the focus of the extracted features to less popular features. These features are supposed to weigh less and therefore be neglected by the targeted system. 
 \end{enumerate}
The difference between the methods of picking the mode elements is a result of the specific evasion attacks, which will be discussed in Section~\ref{ev_attacks}. 
\subsection{Structure Break Algorithm}
\label{mama_att_algo}
\begin{algorithm}[!]
	\caption{Structure Break Attack -  General Algorithm }\label{alg:structure_break}
	\begin{algorithmic}[1]
		\Procedure{Structure Break Attack}{$APK,Mode\_elements,L\_func,P\_func$}
		\State $Manifest,Smali, Layouts \gets depackage(APK)$ \label{lst:line:dpk}
		
		\State $APK\_Tree, Height \gets apk\_structure(Smali)$ \label{lst:line:apk_tree}
		\State $L \gets L\_func(0,Height)$ \label{lst:line:pick_level} 
		\State $P \gets P\_func(0,1)$ \label{lst:line:pick_precent}
		\State $f \gets random(Mode\_elements)$ \label{lst:line:get_elem}
		\State $mkdir(f,APK\_Tree) $ \label{lst:line:cr_elem}
		\State $Roots \gets get\_roots\_random(APK\_Tree,P,L)$ \label{lst:line:get_roots_ratio}
		\ForEach {$r \in  Roots $}\label{lst:line:roots_loop}
\State $r\_new \gets f+r$\label{lst:line:concat}
\State $S\_roots \gets get\_Smali\_files(r)$\label{lst:line:root_get_files} 
\ForEach {$file \in  S\_roots $}\label{lst:line:loop_subs}
\State $file \gets change\_oc(file,r,r\_new)$\label{lst:line:ch_oc_sroots}
\State $ move(r+file,r\_new+file)$
\label{lst:line:move}
\EndFor
\ForEach {$file \in  Manifest,Layouts $}
\label{lst:line:man_lay}

\State $file \gets change\_oc(file,r,r\_new)$
\label{lst:line:ch_man}
\EndFor
\EndFor
		 
		\State $APK \gets Repackage(Manifest,Smali...)$\label{lst:line:repack}
		\State \textbf{return} $APK$\label{lst:line:end}
		\EndProcedure
			\Procedure{change\_oc}{$file,r,r\_new$}{\ForEach {$line \in  file $}\label{lst:line:ch_oc1}
\State $file[line] \gets file[line].replace(r,r\_new)$\label{lst:line:ch_oc2}
\EndFor}
\textbf{return} $file$\label{lst:line:oc_end}
	\EndProcedure
	\end{algorithmic}
\end{algorithm}
Given malicious APKs, the attacker manipulates the structure of each APK towards the picked \textit{mode elements}. An algorithm that implements this manipulation is depicted in Algo.~\ref{alg:structure_break}. The inputs for this algorithm include an APK to manipulate, the \textit{mode elements}, and $L\_func$ and $P\_func$. The last two inputs define two functions. $L\_func$ is a function that given the range of 0 to the directories tree's height (of the application), chooses a specific height. $P\_func$ is a function that given a range of change(the default is [0,1]), chooses a ratio of change (for the specific level chosen by $L\_func$). As there are two options for each function depending on the specific attack variant, they are defined as variables. The algorithm implements the following steps:
(1) The algorithm’s  inputs are an APK file, the 
\textit{Mode\_elements}, $L\_func$ and $P\_func$; (2) Depackage the APK to the Manifest file, Smali code files, and other subordinate files (line~\ref{lst:line:dpk}); (3) Get the structure of the Smali code files as a tree (\textit{APK\_Tree}), and the tree's height (\textit{Height}) (line~\ref{lst:line:apk_tree}); (4) Run $L\_func$ and store the result it \textit{L} (line~\ref{lst:line:pick_level}); (5) Run $L\_func$ and store the result it \textit{P} (line~\ref{lst:line:pick_precent}); (6) Get a random item from the set of Mode\_elements (line~\ref{lst:line:get_elem}) as $f$;
(7) Create a directory whose name is $f$ at the top of \textit{APK\_Tree}, alongside the former root of the tree (line~\ref{lst:line:cr_elem}); (8) According to steps~\ref{lst:line:pick_level}-\ref{lst:line:pick_precent}, get \textit{P} of the directories in level \textit{L} of the \textit{APK\_Tree}. Store them as $Roots$ (line~\ref{lst:line:get_roots_ratio}); (9) For each directory $r$ in the $Roots$ set, run lines~\ref{lst:line:concat}-\ref{lst:line:ch_man} (line~\ref{lst:line:roots_loop}); (10) Concatenate $f$ as a prefix of the former directory name $r$. Call it $r\_new$ (line~\ref{lst:line:concat}); (11) Store all the subdirectories and files of $r$ in \textit{S\_roots} (line~\ref{lst:line:root_get_files}); (12) For each file/directory in $S\_roots$, run lines~\ref{lst:line:ch_oc_sroots}-\ref{lst:line:move} (line~\ref{lst:line:loop_subs}); (13) Run the \textit{change\_oc} on the file/directory, $r$ and $r\_new$ (line~\ref{lst:line:ch_oc_sroots}). The \textit{change\_oc} function (lines~\ref{lst:line:ch_oc1}-\ref{lst:line:ch_oc2}) replaces any occurrence of a line in a file/set of files with a replacement. In this case, it changes the occurrence of $r$ with $r\_new$; (14) Move the file from the previous directory $r$ to the new directory $r\_new$ (line~\ref{lst:line:move}); (15) For each file in the set of Manifest and layout files, run line~\ref{lst:line:ch_man} (line~\ref{lst:line:man_lay}); (16) Run the $change\_oc$ function on the file (line~\ref{lst:line:ch_man}); (17) Repackage the APK (line~\ref{lst:line:repack}), and return it as an output (line~\ref{lst:line:end}).

The attacker creates a new full functional APK, as it changes the structure of the Smali code files, the Manifest file, and the layout files (lines~\ref{lst:line:loop_subs}-\ref{lst:line:ch_man}). As all of these files might include some occurrences of the part that changed (in other words, references to files that moved from their original places), these occurrences need to be changed accordingly. 

A small example of this algorithm is provided for clearance. Let's $APK$ be an APK that includes the following: 
\begin{enumerate}
    \item A Manifest file
    \item 2 Layout Files named $L1$ and $L2$.
    \item A Smali code directory which has the following structure:
    \begin{itemize}
        \item A root directory named $com$.
        \begin{itemize}
        \item A subordinate directory with the name $tb$.
        \begin{itemize}
        \item A Smali code file named $x1.smali$.
        \item A Smali code file named $x2.smali$.
        \end{itemize}
        \item A subordinate directory with the name $xz$.
    \end{itemize}
    \end{itemize}
\end{enumerate}
Let's $mode\_elements={android,java}$, $L\_func=1$, $P\_func=0.5$. For this example, the chosen part for change is the $tb$ folder. The element $f$ is chosen to be $android$.
The structure of the output $APK$ is now:
\begin{enumerate}
    \item A Manifest file
    \item 2 Layout Files named $L1$ and $L2$.
    \item A Smali code directory which has the following structure:
    \begin{itemize}
        \item A root directory named $com$.
        \begin{itemize}
            \item A subordinate directory with the name $xz$.
        \end{itemize}
        \item A root directory named $android$.
        \begin{itemize}
        \item A subordinate directory with the name $tb$.
        \begin{itemize}
        \item A Smali code file named $x1.smali$.
        \item A Smali code file named $x2.smali$.
        \end{itemize}
    \end{itemize}
    \end{itemize}
\end{enumerate}
Each occurrence of the changed part (names of $tb$ directory and its subordinate files) is changed in the Manifest files and $L1$ and $L2$.
\begin{table}
\center
\begin{tabular}{|c|c|}
\hline
Abbreviation & Definition\\
\hline
CV & Cross-Validation\\
\hline
DT & Decision Tree\\
\hline
RF & Random Forest\\
\hline
ML & Machine Learning\\
\hline
ER & Evasion Robustness\\
\hline
DRR & Defense Reciprocal Rate\\
\hline
CFG & Control Flow Graph\\
\hline
APK & Android PacKage\\
\hline
FA & Feature Access\\
\hline
SA & Statistics Access\\
\hline
StB & Structure Break\\
\hline
NN & Nearest Neighbor\\
\hline
TPR & True Positive Rate\\
\hline
\end{tabular}
\caption{Abbreviations table}
\label{abbrev_table}
\end{table}

\subsection{Structure Break Attack Variants}
\label{ev_attacks}
The previous section~\ref{mama_att_algo} described the general StB algorithm. This section describes the actual evasion attacks, which are variants of the StB general algorithm. Each variant is described by its attacker model, $Mode\_elements$, $L\_func$ and $P\_func$.
\begin{enumerate}
    \item \textbf{Random StB Attack:} This attack variant is based on the FA attacker model, as it uses only the knowledge of the feature set. Therefore, a random set of $mode\_elements$ is taken from the feature set. The $L\_func$ an $P\_func$ are random functions. In other words, these functions randomly pick a level in the range of [0,$Height$], and a ratio of change between 0 to 1, respectively. This variant is called Random StB Attack, as the $Mode\_elements$ are chosen randomly.
    \item \textbf{Full Statistical StB Attack:} This attack variant is based on the SA attacker model, which incorporates the knowledge of the feature set and the train data. The $Mode\_elements$ are chosen statistically (for more information, see  Section~\ref{model}). The attacker chooses the element that has the highest values of transitions between families in the benign apps and low values in the malicious apps. Then, it stores it in the $Mode\_elements$. 
     The $L\_func$ and $P\_func$ are not random functions. They are both set to maximize the effect of the change. In other words, $L\_func$ chooses 0 to include the whole smali directory tree in the change of the app. Also, $P\_func$ is set to 1 to include the maximum amount of files and directories in the manipulation. This attack variant is called Full Statistical StB Attack, as it leverages the full knowledge of statistics of the train data.
    \item \textbf{Black Hole Statistical StB Attack:} This attack variant is based on the FA attacker model, which uses only the knowledge of the feature set. The set of $Mode\_elements$ is chosen statistically by the test data only. The attacker chooses the element that holds the lowest values of transitions between families in the test data it obtains (as explained in Section~\ref{model}). The picked element is stored in the $Mode\_elements$. The $L\_func$ and $P\_func$ are identical functions to the functions used in the Full Statistical StB Attack, to maximize the effect of the attack.
    As the idea of this attack variant is to move some of the values to places in the feature vector with no initial positive values (or close to it), which can be referenced as black holes, this variant is called Black hole Statistical StB attack. 
\end{enumerate}

\section{Metrics and Evaluations} \label{eval}
A set of experiments was conducted to evaluate the effects of the evasion attacks of Section~\ref{ev_attacks}. First, Section~\ref{mama_exp_settings} describes the design of the experiments. Then, evaluation metrics are reported in Section~\ref{metrics_names}, followed by their evaluation of the evasion attacks. Section~\ref{former_apps_results} describes the evaluation of the Evasion Robustness (ER) metric. Section~\ref{drr_res} presents the results of the Defense Reciprocal Rate (DRR) metric. At last, Section~\ref{mra_res} describes the evaluation of the Model Reliability. 

\subsection{Experimental Design}
\label{mama_exp_settings}
The experiments include an analysis of an extended set of models of MaMaDroid~\cite{onwuzurike2019mamadroid}. First, the set of distance-based models of MaMaDroid is used (1-NN and 3-NN) and extended by a 5-NN model. Second, the tree-based models are tackled in a parallel way. RF is the only tree-based model that was originally used. As RF is an ensemble of decision trees (DTs), the basic DT model is included. On the other hand, a boosting model (i.e., AdaBoost), which is a more sophisticated ensemble of DTs, is added. In contrast to the RF, where the final decision is based on the decision of each DT independently, in AdaBoost, each DT is aimed to focus on the wrong classification of the prior ones. In total, six models were explored.

For each evasion attack, an experiment was conducted using these six models (1-NN, 3-NN, 5-NN, DT, RF, Adaboost). The experiments were run on an Intel(R) Xeon(R) CPU E5-2683 v4 2.1 GHz with 64 GB RAM with GeForce RTX 2080 TI GPU. The dataset for the experiments consisted of $\sim$73K benign apps from the Google Play Market~\cite{GooglePlay} (obtained from Androzoo~\cite{allix2016androzoo}) and $\sim$6K malicious apps from the Drebin dataset~\cite{ali2016aspectdroid,arp2014drebin,berger2021crystal,frenklach2021android,maiorca2017r,yuan2020byte,zulkifli2018android}. 
To account for variations in the dataset, a 5-fold CV was used. To test the changes in the detection rate of each of the models, different ratios of benign and malicious apps were used during the experiments. The test data without any manipulations will be termed \textit{clean data}. The post-manipulation test data will be referred to as \textit{manipulated data}. To fully evaluate the evasion attacks, an additional experiment on clean data was used as a baseline. To clarify the effect of the amount of benign data on the decision function of each model is tested, the malicious apps are fixed for each experiment, while the benign ratio changes (10\%,20\%,\ldots,100\%).  

\subsection{Metrics}
\label{metrics_names}
\textbf{Evasion Robustness (ER):} To evaluate robustness, the portion of malicious instances which was wrongly classified was computed (similar to the analysis provided in~\cite{tong2017framework}). The TPR of malicious apps was used as this evaluation metric, both for manipulated apps and clean malicious data. 

\textbf{Defense Reciprocal Rank (DRR):} A new metric to evaluate the strength of an evasion attack was presented in~\cite{brama2022evaluation}. The intuition behind the Defense Reciprocal Rank (\textit{DRR}) is that a correct classification is not the only factor of an evasion attack. The confidence of the classifier matters as well. For example, lower confidence may indicate that the attack, though failed to fully deceive the classifier, gained some effect on the classifier and can be more easily improved into a successful attack. Each of the classes (e.g., benign and malicious) is given a rank, based on popularity among the data, target class, and the other class, or any other way of ranking. $\bar{P_i}$ represents the probability assigned to the true class \textit{i}, and $R_i$ is the rank of that class within the ordered predictions list. For the chosen class by the classifier, the $\bar{P_i}$'s range is between 0-1. For the second-best class, $\bar{P_i}$'s range is between 0-0.5, as if it was more than 0.5, it was the chosen class. The other classes will follow a similar rule. The ranks in the case of a binary classification are 1 and 2, where 1 is the highest rank. The ranks in this paper will be picked as 1 to be the right class of the sample, and 2 for the wrong class. Each rank will have a range. The ranges should not overlap as well, as $\bar{P_i}$ will be mapped to a $R_i$'s range. The range of the first rank is between 0.5-1. The second rank will be between 0.33-0.5, and so on. The DRR is calculated in Eq.~\ref{drr_eq}, for a sample $x$:
\begin{equation}
\label{drr_eq}
DRR(x)=\frac{\bar{P_i}}{R_i+1}+\frac{1}{R_i+1}
\end{equation}
The first element maps the range of $\bar{P_i}$ to a range in the size of the $R_i$'s range. The second element adds the lower bound of the $R_i$'s range. A sum of both maps $\bar{P_i}$ to the actual range of $R_i$. Therefore, the overall DRR of a classifier $CL$ on a set of samples $X$ is defined in Eq.~\ref{drr_total_eq}:
\begin{equation}
\label{drr_total_eq}
Overall\_DRR(CL,X)=\frac{\Sigma_{x \in X}DRR(x)}{|X|}
\end{equation}

\textbf{Model Reliability Assessment:} As the process of a binary classification goes, for each test sample, the model outputs a probability of the target class. The second class holds the complement probability. For each test sample, the entropy of the target class and second class is denoted as $s$. In other words, $s=E(P,\overline{P})$, where $E$ symbolizes the entropy function, and $P$,$\overline{P}$ are the probabilities of each class produced by the classifier. To fully evaluate the magnitude of each model, the Shannon Entropy of the probabilities is used, to generate a number that represents the reliability of the model, similarly to the approach suggested in~\cite{nguyen2020ensemble}. The reliability is based on the complement of the average entropy of a test sample. As entropy usually measures uncertainty and disorder, the reliability is calculated as the complement of the average entropy. That way, a higher score is translated as great confidence by the classifier. The reliability is defined with regards to the classifier $CL$, the set of test samples $X$ and the set of entropy $S$ for the test set $X$ using Eq.~\ref{re_eq}:

\begin{equation}
\label{re_eq}
Re(CL,X,S)=1-\frac{S}{|X|}
\end{equation}

The three metrics create an advanced view of the effects of the evasion attacks. The ER metric depicts the gap in detection rates of malicious samples, between clean and manipulated apps. The DRR metric describes the effect the samples have on the certainty of the prediction function of each detection model. The Model Reliability metric concludes the analysis with the overall reliability of each detection model. As will be exampled in the next section, the following scenario can happen between two models: The first model is more affected by the evasion attacks in means of DRR compared to the second. However, the first model is more reliable in facing evasion attacks. Therefore, the DRR and Model Reliability metrics complete one another by means of a thorough analysis of the effects of the evasion attacks. 

\subsection{Results - Evasion Robustness }
\label{former_apps_results}

\begin{figure*}[t!]
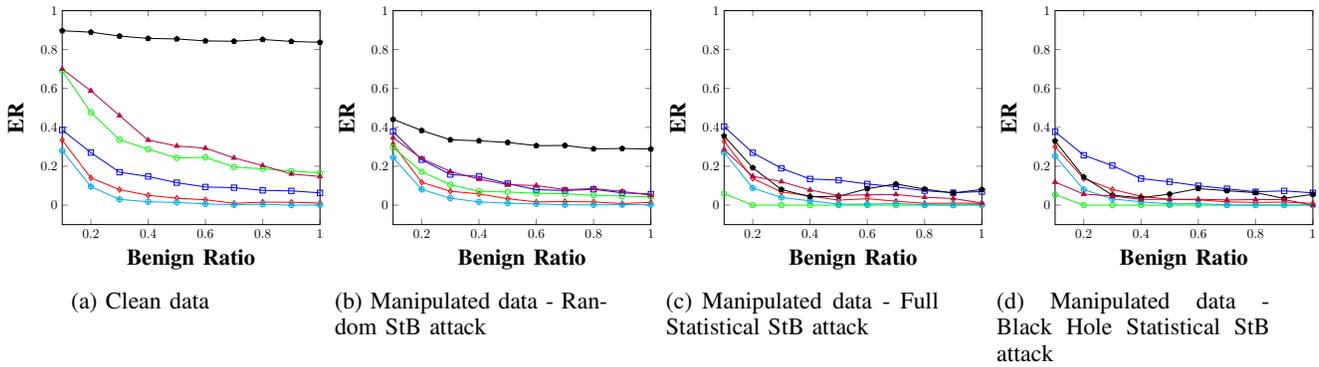

\begin{minipage}[t]{0.2\linewidth}
\centering
\TPRRes{ER}{er_clean_1nn.tex}{er_clean_3nn.tex}{er_clean_rf.tex}{er_clean_5nn.tex}{er_clean_dt.tex}{er_clean_ada.tex}
\subcaption{Clean data}
\label{er_clean}
\end{minipage}
\hspace{15pt}
\begin{minipage}[t]{0.2\linewidth}
\centering
\TPRRes{ER}{er_random_1nn.tex}{er_random_3nn.tex}{er_random_rf.tex}{er_random_5nn.tex}{er_random_dt.tex}{er_random_ada.tex}
\subcaption{Manipulated data - 
Random StB attack}
\label{er_random}
\end{minipage}
\hspace{15pt}
\begin{minipage}[t]{0.2\linewidth}
\centering
\TPRRes{ER}{er_full_1nn.tex}{er_full_3nn.tex}{er_full_rf.tex}{er_full_5nn.tex}{er_full_dt.tex}{er_full_ada.tex}
\subcaption{Manipulated data - 
Full Statistical StB attack}
\label{er_full}
\end{minipage}
\hspace{15pt}
\begin{minipage}[t]{0.2\linewidth}
\centering
\TPRRes{ER}{er_black_1nn.tex}{er_black_3nn.tex}{er_black_rf.tex}{er_black_5nn.tex}{er_black_dt.tex}{er_black_ada.tex}
\subcaption{Manipulated data - 
Black Hole Statistical StB attack}
\label{er_black}
\end{minipage}
\caption{ER results: Distance-based models (1-NN, 3-NN, 5-NN) and Tree-based models (RF, DT, Adaboost). The models were tested with clean data (a), and three types of manipulated data (b-d). The ER represents the detection rate of each model.}
\label{er_eval}
\end{figure*}

\begin{figure*}[t!]
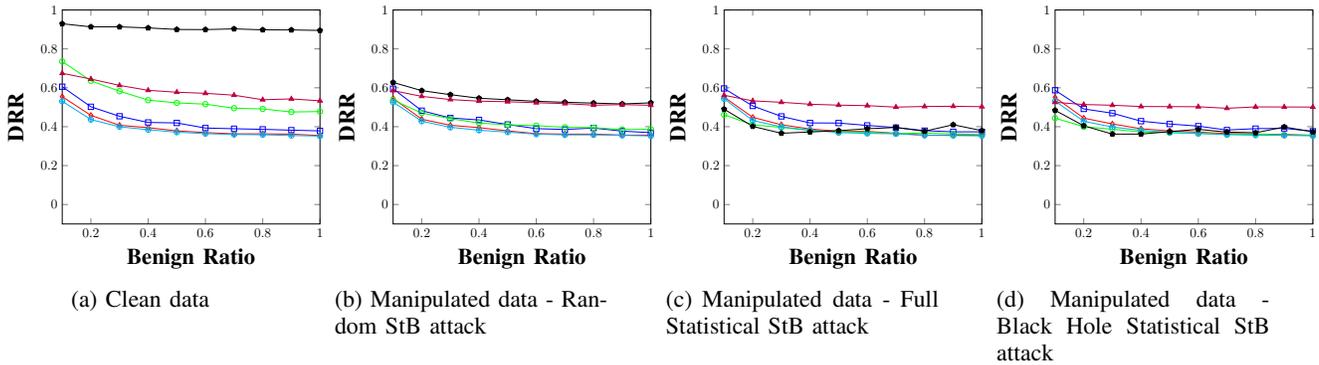

\begin{minipage}[t]{0.2\linewidth}
\centering
\TPRRes{DRR}{drr_clean_1nn.tex}{drr_clean_3nn.tex}{drr_clean_rf.tex}{drr_clean_5nn.tex}{drr_clean_dt.tex}{drr_clean_ada.tex}
\subcaption{Clean data}
\label{drr_clean}
\end{minipage}
\hspace{15pt}
\begin{minipage}[t]{0.2\linewidth}
\centering
\TPRRes{DRR}{drr_random_1nn.tex}{drr_random_3nn.tex}{drr_random_rf.tex}{drr_random_5nn.tex}{drr_random_dt.tex}{drr_random_ada.tex}
\subcaption{Manipulated data - 
Random StB attack}
\label{drr_random}
\end{minipage}
\hspace{15pt}
\begin{minipage}[t]{0.2\linewidth}
\centering
\TPRRes{DRR}{drr_full_1nn.tex}{drr_full_3nn.tex}{drr_full_rf.tex}{drr_full_5nn.tex}{drr_full_dt.tex}{drr_full_ada.tex}
\subcaption{Manipulated data - 
Full Statistical StB attack}
\label{drr_full}
\end{minipage}
\hspace{15pt}
\begin{minipage}[t]{0.2\linewidth}
\centering
\TPRRes{DRR}{drr_black_1nn.tex}{drr_black_3nn.tex}{drr_black_rf.tex}{drr_black_5nn.tex}{drr_black_dt.tex}{drr_black_ada.tex}
\subcaption{Manipulated data - 
Black Hole Statistical StB attack}
\label{drr_black}
\end{minipage}
\caption{DRR results: Distance-based models (1-NN, 3-NN, 5-NN) and Tree-based models (RF, DT, Adaboost). The models were tested with clean data (a), and three types of manipulated data (b-d). The DRR represents the DRR results of each model.}
\label{drr_eval}
\end{figure*}

\begin{figure*}[t!]
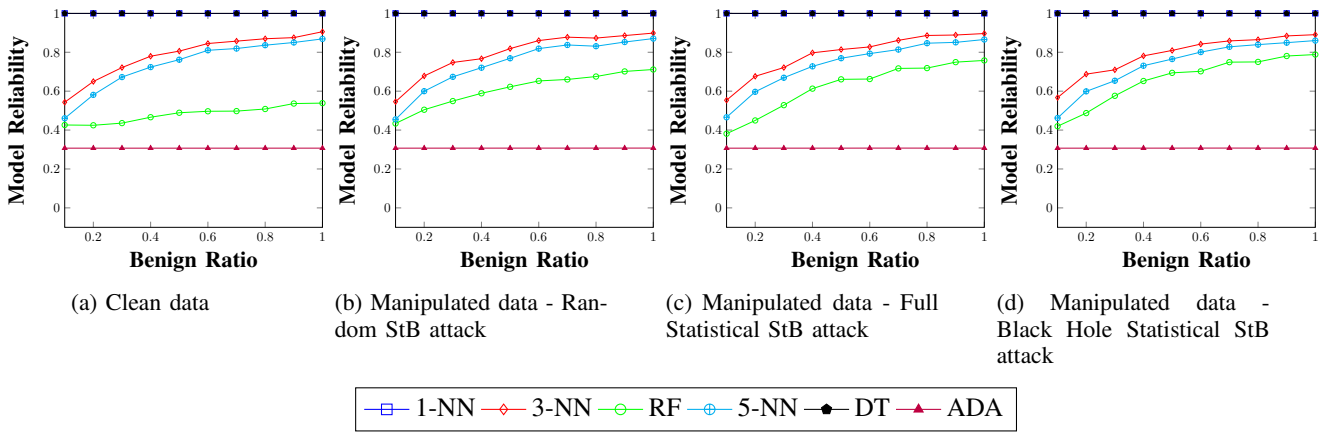

\begin{minipage}[t]{0.2\linewidth}
\centering
\TPRRes{Model Reliability}{mr_clean_1nn.tex}{mr_clean_3nn.tex}{mr_clean_rf.tex}{mr_clean_5nn.tex}{mr_clean_dt.tex}{mr_clean_ada.tex}
\subcaption{Clean data}
\label{mr_clean}
\end{minipage}
\hspace{15pt}
\begin{minipage}[t]{0.2\linewidth}
\centering
\TPRRes{Model Reliability}{mr_random_1nn.tex}{mr_random_3nn.tex}{mr_random_rf.tex}{mr_random_5nn.tex}{mr_random_dt.tex}{mr_random_ada.tex}
\subcaption{Manipulated data - 
Random StB attack}
\label{mr_random}
\end{minipage}
\hspace{15pt}
\begin{minipage}[t]{0.2\linewidth}
\centering
\TPRRes{Model Reliability}{mr_full_1nn.tex}{mr_full_3nn.tex}{mr_full_rf.tex}{mr_full_5nn.tex}{mr_full_dt.tex}{mr_full_ada.tex}
\subcaption{Manipulated data - 
Full Statistical StB attack}
\label{mr_full}
\end{minipage}
\hspace{15pt}
\begin{minipage}[t]{0.2\linewidth}
\centering
\TPRRes{Model Reliability}{mr_black_1nn.tex}{mr_black_3nn.tex}{mr_black_rf.tex}{mr_black_5nn.tex}{mr_black_dt.tex}{mr_black_ada.tex}
\subcaption{Manipulated data - 
Black Hole Statistical StB attack}
\label{mr_black}
\end{minipage}
\begin{center}
    \ref{barLagendName}
\end{center}
\caption{Model Reliability results: Distance-based models (1-NN, 3-NN, 5-NN) and Tree-based models (RF, DT, Adaboost). The models were tested with clean data (a), and three types of manipulated data (b-d). The Model Reliability of each model is presented.}
\label{mr_eval}
\end{figure*}

The first evaluation analyzes the effect of the benign ratio on the ER.
The ER metric was computed using the TPR of the malicious apps as the original detection rate was computed. Naturally, as one can expect, as the ratio of benign samples in the train set increases, the classifier will tend to focus on these samples, aiming to minimize the loss function, and the TPR will monotonically decrease. 
The results of the ER are split into four parts\footnote{extended version of these results and the results of the evasion attacks can be found in Github:https://github.com/harelber/The-Black-Holes-of-Markov-Chains.}, depicted in Fig.~\ref{er_eval}. First, the ER of the clean data is presented as a baseline in Fig.~\ref{er_clean}. Second, the ER of the Random StB attacks is described in Fig.~\ref{er_random}. Then, the ER of the Full Statistical StB attack is presented  in Fig.~\ref{er_full}. At last, the ER of the Black Hole Statistical StB attack is depicted in Fig.~\ref{er_black}.
\begin{enumerate}[a)]
    \item\textbf{Clean data:} The results of the ER of the clean data can be seen in Fig.~\ref{er_clean}. It can be seen that the more benign apps used, the fewer malicious apps are detected by each one of the models. The 1-NN and 3-NN models' starting points at 10\% of the benign apps (blue and red lines) are lower than the RF (green line), by $\sim$30\%. Their ending point with 100\% of the benign apps reaches almost 0\%. In other words, the RF's detection rate decreases by $\sim50$\%, and the KNNs by $\sim40$\%. The DT's (black line) starting point at 10\% of the benign apps, is a detection rate of $\sim$90\%. Moreover, this model sustains a stable detection rate. The Adaboost model (red line) shows similar detection rates to the RF, 1-NN and 3-NN, with a slight superiority over the RF model. The 5-NN model (cyan line) is less accurate than all models. It seems that the DT is the best model out of the six.
    \item \textbf{Random StB attack:} \label{random_stb_results}
The results of the ER evaluation of the manipulated data using the Random StB attack are depicted in Fig.~\ref{er_random}. The distance-based models (1-NN, 3-NN, and 5-NN) show an interesting behavior. They have a similar detection rate along the interval of the benign ratio between the clean data and the manipulated data. In other words, the attack did not have a great effect on them. The RF model suffers from a great loss of $\sim$40\% along the interval. In other words, the attack affected the RF model dramatically. This is a change of course, as in the case of the clean data, the RF was more accurate than the 1-NN and 3-NN. The DT and Adaboost models have a similar distribution as with the clean data. However, the starting and ending points of each model in both of them are lower than the correlative points in the clean data's results. The DT model is stable and has the most effective results against the Random StB attack. However, it has an ER of less than 50\% with only 10\% of the benign apps. It falls under 40\% with 100\% of the benign apps. The Adaboost model is now as effective as the 1-NN model.

\item \textbf{Full Statistical StB attack:} \label{fs_stb_results}
In this case, which is depicted in Fig.~\ref{er_full}, the ER of the manipulated data by the Full Statistical StB attack is presented. The distance-based models stay the same as in previous cases. Their ER is stable, despite their low rates. The RF model continues its decrease in ER, and in this case - to $\sim$0\% on most of the interval. The DT and Adaboost models decrease their ER as well, by 20\%-30\%.

\item \textbf{Black Hole Statistical StB attack:} \label{bh_stb_results}
The ER results of the manipulated data by the Black Hole Statistical StB attack are depicted in Fig.~\ref{er_black}. In this case, the distance-based models stay the same as before. The ER of the tree-based models (DT, RF, Adaboost) continue their decrease in ER, and in this case - less than 10\% on most of the interval. 

\end{enumerate}

The distance-based models show poor results with the clean data and against the evasion attacks as well. In comparison, the tree-based models show high ER with clean data. Also, the tree-based models were proven to be more susceptible to attacks than the distance-based models. An explanation for these findings is based on the split nodes function of the tree models. The node split function is effective in means of identification of malicious activity, without any manipulation. The more features it processes, the more accurate it becomes. This is true for the DT model, as the amount of benign data did not dramatically affect its detection rate. However, splitting the data between multiple weaker modules, like exampled in the Adaboost and RF models, results in a decrease in the ER. Also, the tree-based models are based on features that are often manipulated by the attack. Therefore, the ER changes according to the attacks. However, as most of the features stay similar to the clean data, the distance-based models show a similar detection rate between the baseline and the evasion attacks. 

\subsection{Results - Defence Reciprocal Rate}
\label{drr_res}
The second evaluation is of the DRR. The DRR measures how effective an evasion attack is. A higher DRR means a stronger classifier. In addition to the ER, this metric reflects the gap between the predictions on clean data and the predictions on evasion attacks. The ranks chosen for this experiment were 1 for malicious and 2 for benign, as the test data consisted of only malicious data. Therefore, the rank of 1 depicts the correct label, and a rank of 2 for a mistake. In this section, a review of the results of the models tested on clean data, and the three evasion attacks is presented. The clean data results are depicted in Fig.~\ref{drr_clean}. The DRR of the evasion attacks are presented in Fig.~\ref{drr_random}-\ref{drr_black}.

\begin{enumerate}
    \item\textbf{Clean data: }The results of the DRR evaluation of the clean data are depicted in Fig.~\ref{drr_clean}. As with the case of the ER, the tree-based models overcome the distance-based models, considering data without any evasion attack. Also, the DT model holds the highest DRR, which depicts the most resilient model. Its DRR is more than 90\% along the interval.
    \item \textbf{Random StB attack:} In Fig.~\ref{drr_random}, the DRR evaluation of the manipulated data using the Random StB attack is depicted. The DRR of the distance-based models stays the same as in the clean data. The RF model's DRR suffers from a decrease of $\sim$20\% in comparison to the clean data. The DT model's DRR decreases by $\sim$30\% in comparison to the clean data. The decrease for the Adaboost model is $\sim$20\% in comparison to the clean data. The DT and Adaboost models show close results along the interval.
    \item \textbf{Full Statistical StB attack:} In Fig.~\ref{drr_full}, the DRR of the manipulated data by the Full Statistical StB attack is presented. The distance-based models stay with a steady DRR as before. The Adaboost model shows surprising results, similar to the DRR of the Random StB attack. In other words, the attacks had a similar effect on the DRR of the Adaboost model, in contrast to the DT model, which decreased by an additional 10\% in comparison to the Random StB attack. This might raise an alert that the DT model is less reliable than the Adaboost model. However, the next section on the Model Reliability metric will answer this question. The RF model suffers from another DRR decrease, of $\sim$10\% in comparison to the Random StB attack.
    \item \textbf{Black Hole Statistical StB attack:} The DRR evaluation of the manipulated data by the Black Hole Statistical StB attack is depicted in Fig.~\ref{drr_black}. The distance-based models, the DT and Adaboost models present identical DRR to the Full Statistical StB attack. The starting point of the RF is lower than its correlative starting point in the Full Statistical StB attack.
    \end{enumerate}

 The distance-based models showed similar DRRs in each case explored, as in the ER metric. The evasion attacks had a similar effect on the DRR of the Adaboost model, in contrast to the DT model, which decreased by 50\% in the case of the Random StB attack, and an additional 20\% in the Full and Black Hole Statistical attacks. These findings raise the need to investigate if the Adaboost model is more effective than the DT model and if the Adaboost model is more reliable. The next section on the Model Reliability metric will answer these inquiries.

\subsection{Results - Model Reliability Assessment}
\label{mra_res}

The reliability of each model was inspected, to understand the confidence of each model on the predictions. High reliability means that the gap between the two classes is great. A low value means that the classes are close to being indistinguishable. This does not influence the ER, as a classifier with a low detection rate can be "confident" in its predictions. Also, a "lucky" classifier can have a high detection rate, but the probabilities of each class may be close. The assessment was done on four cases - clean data and the manipulated data by the three evasion attacks - to compare the changes the attacks have on the classifiers. The assessment is depicted in Fig.~\ref{mr_eval}. As the reliability assessment results are similar between the four cases - clean data and manipulated data - they will be discussed without specific references to each one of the subfigures. 

It can be seen that the 1NN model is reliable, as its reliability is 1 on the whole interval, for each one of the cases. The reliability of the DT model is also found to be very high, as its reliability receives the value of 1 along whole the interval for each one of the cases. The reliability of the Adaboost model was found to be a constant value of 0.3 along the interval, for each one of the cases.
The Model reliability rates of the other models - 3NN, 5NN, and RF are increasing throughout the interval. The increasing reliability of the 3NN and 5NN models is similar in the four cases. The RF model is found to be more reliable in the cases of evasion attacks than the clean data. In other words, it is more "confident" in its labels of the evasion attacks than the clean data. 

Both the 1NN and DT models show a constant high value of model reliability, with a value of 1. In contrast, the Adaboost model has a constant value of 0.3. As these phenomena were surprising, a close examination of the probabilities of each class for these models was done. It was found that for the 1NN and DT models, the probabilities are binary for each sample, without any regard to the benign ratio of the apps. For the Adaboost model, both probabilities on each sample were close to 0.5. Therefore, the entropy is high in the results of the Adaboost model, and low in the results of the 1NN and DT models, which outcomes in correlative complement Model Reliability rates. Therefore, it was found that although the Adaboost model had better DRR than the DT model against the evasion attacks, its reliability is far worse. The DT model was found as the most promising model out of the six, based on the combined insights of the ER, DRR, and Model Reliability metrics.

Four important insights are derived from the combination of the results of the ER, DRR, and Model Reliability: 
\begin{enumerate}
    \item The most promising model tested on the clean data and the evasion attacks is DT. However, this model is susceptible to evasion attacks.
    \item The distance-based models were not much affected by the evasion attacks. However, their detection rate is low in the baseline and any evasion attack.  
    \item The ER and DRR results proved that the Random StB attack is less effective than the Full Statistical and the Black Hole Statistical attacks. For example, for the DT model, the DRR of the clean data was 80\%-90\%. On the Random StB attack, the DRR was 50\%-60\%. However, the DRR decreased to 40\% and less in the cases of the Statistical StB attack and Black Hole Statistical attack. Also, the ER supports this conclusion. This insight is pretty intuitive as a random attack is not specifically targeting the data the model is built upon. In comparison, the Full Statistical StB attack requires the train data to generate the attack. However, it was important to establish this insight by viewing the results. In addition, the Black Hole Statistical StB attack got similar results to the Full Statistical StB attack but needs less information - only the test data. Therefore, it can be considered the best attack out of the three. 
    \item A lower DRR does not necessarily means a less effective model. As proven, the DT model had a lower DRR in the case of the Full and Black Hole Statistical StB attacks than the Adaboost model, which had a similar DRR for each attack. However, the ER and Model Reliability showed the full picture, in which the Adaboost model is less reliable and robust than the DT model.
    
\end{enumerate} 
A more robust version of MaMaDroid can be suggested, due to the effect of the ratio of benign apps on the detection of malicious apps and evasion attacks. This version should use the most effective model that was found - DT. However, using the existing DT model is not enough, as it is susceptible to StB attacks. The next section will demonstrate a suitable solution for the holes in the model.

\section{MaMaDroid2.0}
\label{mama2}
As suggested in the previous section, DT was found to be the most effective and reliable model for MaMaDroid. However, the DT model was found to be vulnerable to StB attacks. The distance-based models were found to be robust against the evasion attacks, but their original detection rates were low. A stronger version of MaMaDroid can be suggested due to these findings. As even the strongest model, DT was evaded by the StB attacks, there is still room for improvement. An extension of the feature set might help to mitigate the attacks, thus creating a hybrid approach. 

One of the influential works on Android malware detection is Drebin~\cite{arp2014drebin}. This work laid the foundations for ML Android malware detection which is based on static analysis. This system is based on eight feature sets. For the extension of
MaMadroid, one of the feature sets was picked - the required permissions set. Permissions are a great candidate for a feature set of Android malware detection systems over the years, as a full feature set of a detection machine or a part of it (i.e.~\cite{arora2019permpair,aswini2014droid,ding2021hybrid,enck2009lightweight,li2018significant,pierazzi2020problemspace,sanz2013puma,talha2015apk}). Therefore, the required permissions set was picked to enhance MaMaDroid. It will be termed as \textit{permission set}.

An analysis of the permission set of the dataset (the same dataset as Section~\ref{eval}) resulted in several insights. First, using the full permission set from the dataset will increase the size of the feature set. Second, the less frequent permissions will not affect the detection process as their weight will be low. Moreover, an analysis of the rare permissions found that they were custom permissions~\cite{Android_permissions_cust} used on a few apps. Therefore, only the frequently requested permissions in the dataset were included in the permission set, in over 10\% of the data, both benign and malicious\footnote{To assess the applicability of the permission set for other datasets, a subset of 4K malicious apps and 2K benign apps of MaMaDroid1.0 dataset was analyzed as well. The permission set that was extracted had an intersection of more than 87\% with the permission set that was used in this paper. This proved that the permission set, though needs to be updated from time to time, can be useful for the classification of other datasets as well.}. 

MaMaDroid 2.0, the enhanced version of MaMaDroid, follows the following steps:

\begin{enumerate}
    \item Run MaMaDroid1.0's feature extraction on the app to get MaMaDroid1.0's feature set.
    \item Extract the permission set from each app by exploring its Manifest file. Filter the less frequently requested permissions.
    \item Merge the two feature sets.
    \item Run the model on the merged feature set and get a classification of the app.
\end{enumerate}

The following sections include the experimental design of MaMaDroid2.0 (Section~\ref{2_set}), and the results (Section~\ref{2_res}) 
\subsection{Experimental Design}
\label{2_set}
Several experiments were done to assess the power of MaMaDroid2.0. The dataset for these sets of experiments consisted of the dataset from Section~\ref{eval}, using 100\% of the benign apps. As will be seen in the results, using 100\% of the benign apps does not damage the effectiveness of MaMaDroid2.0, concerning the  classification of benign and malicious apps. This set of experiments tested the effectiveness of the new detection machine, both on clean data and manipulated data. It was done to see if the addition of features damaged the correct classification of benign or malicious data. 

In each experiment, the DT model was tested, which was proven to be the best model in Section~\ref{eval}. Each test case consisted of benign and malicious data, where the malicious data changed according to the specific case. The model was tested in terms of f1-score and recall, to test its correct detection as a whole, and specifically the malicious samples. As most of the test set is benign, it is pretty easy to detect benign apps as benign, using the trivial labeling of the whole test data as benign. Therefore, the recall metric was used, to identify the correct classification of malicious samples. The F1 metric was used to see the effect of the precision and recall together, so as to see the false positives rate on the new machine along with the recall. 

Furthermore, as MaMaDroid2.0 incorporates the permission set as part of its feature set (as used in Drebin~\cite{arp2014drebin}),  attacks against Drebin~\cite{berger2021crystal} were tested as well, to see if the new machine is strong against these attacks as well.

Each experiment is described by the test data and feature set, as the train data stays the same. The settings are described in Table~\ref{exp_table}.
\begin{table}
\begin{tabular}{|c|c|c|c|c|c|}
\hline
Exp & Clean&Manipulated&StB&MB~\cite{berger2021crystal}&Feature Set\\
\hline
1 & \checkmark&\checkmark&\checkmark&&Base\\
\hline
2 & \checkmark&\checkmark&\checkmark&&Ext\\
\hline
3 & \checkmark&\checkmark&\checkmark&&Perm\\
\hline
4 & \checkmark&\checkmark&&\checkmark&Ext\\
\hline
5 & \checkmark&\checkmark&&\checkmark&Perm\\
\hline
\end{tabular}
\caption{MaMaDroid2.0 experiments settings. Each experiment is described by the data it incorporates - clean and manipulated data, the evasion attack that created the manipulated data (StB/MB~\cite{berger2021crystal}). Also, the and the feature set that was used is presented - the basic MaMaDroid1.0 feature set (Base), the extended feature set of MaMaDroid2.0 (Ext), and the permission set (Perm).}
\label{exp_table}
\end{table}
The results of these five experiments are presented in Section~\ref{2_res}.
\subsection{Results - MaMaDroid2.0}
\label{2_res}
The results of the first three experiments of MaMaDroid2.0 are presented in Table~\ref{res_mama2_stb}. It can be seen that the recall of the baseline of DT, as was seen in the results of the former experiments (Section~\ref{eval}), is 0.84 using clean data. Additionally, the recall of the StB attacks is lower than 0.3. However, adding the permission set thus using the enhanced feature set, raises the recall of each case to 0.9 or more. The f1-score of each case using this extended feature set is more than 0.96. In comparison, in the baseline, the f1-score is 0.89 with the clean data and less than 0.66 with the manipulated data. Using only the permission set achieves a close recall and f1-score rates to the enhanced feature set of MaMaDroid2.0. 
\begin{table}
\begin{tabular}{|l|c|c|c|c|c|c|c|c|}
\hline
\multicolumn{1}{|c|}{Cases} & \multicolumn{2}{c|}{Clean} & \multicolumn{2}{c|}{Random}
& \multicolumn{2}{c|}{Full}
& \multicolumn{2}{c|}{Black}\\
\cline{2-9}
\multicolumn{1}{|c|}{FS} & R & F1 & R & F1& R & F1 & R & F1\\
\hline
Base   & 0.84    &0.89&0.27&0.65&0.07
&0.52&0.05&0.51\\
\hline
Ext&   0.96  & 0.97&0.92&0.97 &0.9
&0.96&0.9&0.96  \\
\hline
Perm & 0.91&0.95&0.92 &0.95
&0.92&0.95&0.92&0.96\\
\hline
\end{tabular}
\caption{MaMaDroid 2.0 Recall (R) and F1 metrics for the experiments with the StB attacks. The Feature sets (FS) include the original MaMaDroid features (Base), extended feature set (Ext), and permission set only (Perm). The cases inspected are clean data (Clean), Random StB attack (Random), Full Statistical StB attack (Full), and Black Hole Statistical StB attack (Black).}
\label{res_mama2_stb}
\end{table}

These results may raise the question: What is the importance of the MaMaDroid1.0 feature set if the permission set alone can be used as a standalone to identify evasion attacks? The answer lies in Table~\ref{res_mama2_mb}. The MB attacks show a great impact on a model that is based on the permission set. The highest recall using the permission set only is against MB1, with a value of 0.49. The other MB attacks achieve recall values of 0. In comparison, the enhanced feature set achieves high recall values, at least 0.97, for each case - clean data and manipulated data. This proves that a model that analyzes only the permission set cannot identify evasion attacks that manipulate permissions. In contrast, MaMaDroid2.0 succeeds in this task. 

In conclusion, it was proven that the enhanced feature set of MaMaDroid2.0 creates a stronger model. This model was proved to identify attacks against both MaMaDroid and Drebin alike. 

\begin{table}
\begin{tabular}{|l|c|c|c|c|c|c|c|c|}
\hline
\multicolumn{1}{|c|}{Cases} & \multicolumn{2}{c|}{Clean} & \multicolumn{2}{c|}{MB1}
& \multicolumn{2}{c|}{MB2}
& \multicolumn{2}{c|}{MB3}\\
\cline{2-9}
\multicolumn{1}{|c|}{FS} & R & F1 & R & F1& R & F1 & R & F1\\
\hline
Ext   & 0.97    &0.93&0.98&0.92&0.98
&0.92&0.97&0.91\\
\hline
Perm&   0.92  & 0.95&0.49&0.79 &0
&0.47&0&0.46  \\
\hline
\end{tabular}
\caption{MaMaDroid 2.0 Recall (R) and F1 metrics for the experiments with the MB attacks. The Feature sets (FS) include the extended feature set (Ext) and permission set only (Perm). The cases inspected are clean data (Clean), MB1 attack (MB1), MB2 attack (MB2), and MB3 attack (MB3).}
\label{res_mama2_mb}
\end{table}

\section{Discussion and Conclusion}
\label{diss}
The original MaMaDroid was based on a specific ratio between benign and malicious applications. The original ratios were between 50\%-50\%, or with a tendency toward the malicious class. As a recent study suggested~\cite{pendlebury2018tesseract}, the realistic ratio in the world between benign and malicious apps is leaning toward the benign class. More specifically, it is 90\%-10\% between benign and malicious apps. As a consequence, the number of benign apps was tested in this research to derive their influence on the ER. As was seen, the ratio did have an impact on each one of the models that were tested. In addition, the StB evasion attacks decreased the tree-based models' ERs. The distance-based models proved to be strong against the attacks, but their detection rates were already low on the clean data (and stayed there on the manipulated data by attacks). The DT model was found to be the best model by means of ER. It was also found to be very reliable, even in the most challenging ratio of more than 90\% of the dataset being benign apps. For all of the above, it was suggested to replace the former models of MaMaDroid with a DT. In addition, the merged feature set upgraded the total effectiveness of the model. Adding the enumeration of permission requests, which are a small amount of data, aids the identification of malicious activities. Other additions to the feature set may help in this task, such as suspicious API calls or intents. These directions are left as future work.

Most mitigation techniques to evasion attacks try to find the trail, the attacks leave and identify them. With other attacks, the effect may be eliminated by preprocessing. For example, adding no-operation calls to CFG based detection machines, like in~\cite{chen2018android}. In regards to the StB attacks, the flow of the attack does not add actual code to the app, just a replacement of references to classes names, and moving parts of the application between places. Moreover, in the Full/Black Hole Statistical StB attacks, there is only an addition of a new root directory to the app. A mitigation technique that might succeed in some of the cases is looking for duplication of a name of a directory. For example, if the former root directory of the app is "android" and after the evasion attack now the root directory is also "android", there will be two android directories, one below the other. A quick check for such a case will result in an alert for the StB case. However, not every app's root directory is similar to the new root directory which is picked by the attacker. Therefore, the StB attacks are still a threat to malware detection machines that are based solely on CFG.

MaMaDroid is an excellent example of using the CFG to identify behaviors of benign and malicious APKs. In general, a CFG of a software tries to organize the order of commands the software runs. A malware detection machine that is based on a CFG tries to find the differences between the order of commands in malicious and benign apps to address the task of the classification of new samples. As the experiments in this study showed, a shift in the order of commands can cause a miss-classification. For example, the Black Hole StB variant demonstrates the effect of moving some of the commands from the malicious order of commands to an unknown order of commands. In other words, the order turns to chaos. The different variants modify the order of commands in an app, which results in the alteration of its CFG. When the new CFG is inspected by the detection machine, it is puzzled. This course of action can be generalized to other domains as well. Therefore, the CFG-based detection machines are now faced with the following challenge: Are they secure enough against the ominous chaos that will arrive? It seems that an extension of the feature set or a hybrid approach may aid in dealing with this issue. The combination of different features seems a promising step towards more accurate solutions against multiple types of malware that are exploiting different vulnerabilities.

\section{Acknowledgment}
This work was supported by the Ariel Cyber Innovation Center in conjunction with the Israel National Cyber Directorate of the Prime Minister's Office. The authors will like to thank Dr. Lucky Onwuzurike for his technical aid with MaMaDroid.

\bibliographystyle{plain}
\bibliography{bibliography}

\end{document}

%% file: macros.tex
\newcommand{\TPRRes}[7]{

\begin{tikzpicture}[scale=0.5, transform shape]
\begin{axis}[legend to name=barLagendName,legend columns=-1,
xlabel={\LARGE \textbf{Benign Ratio}},
  ylabel={\LARGE \textbf{#1}},
   xmin=0.1, xmax=1,
    ymin=-0.1, ymax=1,
scatter/classes={
1nn={mark=square,blue},
3nn={mark=diamond,red},
rf={mark=o,green},
5nn={mark=oplus,cyan},
dt={mark=pentagon*,black},
ada={purple,mark=triangle*}
}
]
\addplot[scatter, blue, mark=square] table[x=x, y=y]{#2};
\addplot[scatter, red, mark=diamond] table[x=x, y=y]{#3};
\addplot [scatter, green,
mark=o]table[x=x, y=y]{#4};
\addplot [scatter, cyan,
mark=oplus]table [x=x, y=y]{#5};
\addplot [scatter, black,
mark=pentagon*] table[x=x, y=y]{#6};
\addplot [scatter, purple,
mark=triangle*] table[x=x, y=y]{#7};
\legend{\normalsize 1-NN,3-NN,RF,5-NN,DT,ADA}

\end{axis}
\end{tikzpicture}

}

%% file: usenix2019_v3.1.bbl
\begin{thebibliography}{10}

\bibitem{GooglePlay}
Google.
\newblock {GooglePlay app market}.
\newblock GooglePlay website, 2008.
\newblock \url{https://play.google.com/store/apps/}.

\bibitem{covid-report}
BUSINESS WIRE.
\newblock A year of lockdown sees a surge in mobile malware targeting banking,
  billing and covid-19 vaccines.

\bibitem{aafer2013droidapiminer}
Yousra Aafer, Wenliang Du, and Heng Yin.
\newblock Droidapiminer: Mining api-level features for robust malware detection
  in android.
\newblock In {\em International Conference on Security and Privacy in
  Communication Systems}, pages 86--103. Springer, 2013.

\bibitem{arp2014drebin}
Daniel Arp, Michael Spreitzenbarth, Malte Hubner, Hugo Gascon, Konrad Rieck,
  and CERT Siemens.
\newblock Drebin: Effective and explainable detection of android malware in
  your pocket.
\newblock In {\em Ndss}, volume~14, pages 23--26, 2014.

\bibitem{berger2021crystal}
Harel Berger, Chen Hajaj, Enrico Mariconti, and Amit Dvir.
\newblock Crystal ball: From innovative attacks to attack effectiveness
  classifier.
\newblock {\em IEEE Access}, 2021.

\bibitem{cai2018towards}
Haipeng Cai and John Jenkins.
\newblock Towards sustainable android malware detection.
\newblock In {\em Proceedings of the 40th International Conference on Software
  Engineering: Companion Proceedings}, pages 350--351. ACM, 2018.

\bibitem{chen2016stormdroid}
Sen Chen, Minhui Xue, Zhushou Tang, Lihua Xu, and Haojin Zhu.
\newblock Stormdroid: A streaminglized machine learning-based system for
  detecting android malware.
\newblock In {\em Proceedings of the 11th ACM on Asia Conference on Computer
  and Communications Security}, pages 377--388. ACM, 2016.

\bibitem{dini2012madam}
Gianluca Dini, Fabio Martinelli, Andrea Saracino, and Daniele Sgandurra.
\newblock Madam: a multi-level anomaly detector for android malware.
\newblock In {\em International Conference on Mathematical Methods, Models, and
  Architectures for Computer Network Security}, pages 240--253. Springer, 2012.

\bibitem{huynh2017new}
Ngoc~Anh Huynh, Wee~Keong Ng, and Kanishka Ariyapala.
\newblock A new adaptive learning algorithm and its application to online
  malware detection.
\newblock In {\em International Conference on Discovery Science}, pages 18--32.
  Springer, 2017.

\bibitem{enck2009lightweight}
William Enck, Machigar Ongtang, and Patrick McDaniel.
\newblock On lightweight mobile phone application certification.
\newblock In {\em Proceedings of the 16th ACM conference on Computer and
  Communications Security}, pages 235--245. ACM, 2009.

\bibitem{onwuzurike2019mamadroid}
Lucky Onwuzurike, Enrico Mariconti, Panagiotis Andriotis, Emiliano~De
  Cristofaro, Gordon Ross, and Gianluca Stringhini.
\newblock Mamadroid: Detecting android malware by building markov chains of
  behavioral models.
\newblock In {\em Annual Symposium on Network and Distributed System Security},
  2017.

\bibitem{shabtai2012andromaly}
Asaf Shabtai, Uri Kanonov, Yuval Elovici, Chanan Glezer, and Yael Weiss.
\newblock ``andromaly'': a behavioral malware detection framework for android
  devices.
\newblock {\em Journal of Intelligent Information Systems}, 38(1):161--190,
  2012.

\bibitem{shabtai2009detection}
Asaf Shabtai, Robert Moskovitch, Yuval Elovici, and Chanan Glezer.
\newblock Detection of malicious code by applying machine learning classifiers
  on static features: A state-of-the-art survey.
\newblock {\em Information Security Technical Report}, 14(1):16--29, 2009.

\bibitem{shabtai2014mobile}
Asaf Shabtai, Lena Tenenboim-Chekina, Dudu Mimran, Lior Rokach, Bracha Shapira,
  and Yuval Elovici.
\newblock Mobile malware detection through analysis of deviations in
  application network behavior.
\newblock {\em Computers \& Security}, 43:1--18, 2014.

\bibitem{sun2016sigpid}
Lichao Sun, Zhiqiang Li, Qiben Yan, Witawas Srisa-an, and Yu~Pan.
\newblock Sigpid: significant permission identification for android malware
  detection.
\newblock In {\em 2016 11th International Conference on Malicious and Unwanted
  Software}, pages 1--8. IEEE, 2016.

\bibitem{talha2015apk}
Kabakus~Abdullah Talha, Dogru~Ibrahim Alper, and Cetin Aydin.
\newblock Apk auditor: Permission-based android malware detection system.
\newblock {\em Digital Investigation}, 13:1--14, 2015.

\bibitem{treadwell2009heuristic}
Scott Treadwell and Mian Zhou.
\newblock A heuristic approach for detection of obfuscated malware.
\newblock In {\em 2009 IEEE International Conference on Intelligence and
  Security Informatics}, pages 291--299. IEEE, 2009.

\bibitem{venugopal2008efficient}
Deepak Venugopal and Guoning Hu.
\newblock Efficient signature based malware detection on mobile devices.
\newblock {\em Mobile Information Systems}, 4(1):33--49, 2008.

\bibitem{wang2014exploring}
Wei Wang, Xing Wang, Dawei Feng, Jiqiang Liu, Zhen Han, and Xiangliang Zhang.
\newblock Exploring permission-induced risk in android applications for
  malicious application detection.
\newblock {\em IEEE Transactions on Information Forensics and Security},
  9(11):1869--1882, 2014.

\bibitem{wu2012droidmat}
Dong-Jie Wu, Ching-Hao Mao, Te-En Wei, Hahn-Ming Lee, and Kuo-Ping Wu.
\newblock Droidmat: Android malware detection through manifest and api calls
  tracing.
\newblock In {\em 2012 Seventh Asia Joint Conference on Information Security},
  pages 62--69. IEEE, 2012.

\bibitem{xu2013permlyzer}
Wei Xu, Fangfang Zhang, and Sencun Zhu.
\newblock Permlyzer: Analyzing permission usage in android applications.
\newblock In {\em 2013 IEEE 24th International Symposium on Software
  Reliability Engineering}, pages 400--410. IEEE, 2013.

\bibitem{maggi2013two}
Federico Maggi, Alessandro Frossi, Stefano Zanero, Gianluca Stringhini, Brett
  Stone-Gross, Christopher Kruegel, and Giovanni Vigna.
\newblock Two years of short urls internet measurement: security threats and
  countermeasures.
\newblock In {\em proceedings of the 22nd international conference on World
  Wide Web}, pages 861--872, 2013.

\bibitem{rahbarinia2017exploring}
Babak Rahbarinia, Marco Balduzzi, and Roberto Perdisci.
\newblock Exploring the long tail of (malicious) software downloads.
\newblock In {\em 2017 47th Annual IEEE/IFIP International Conference on
  Dependable Systems and Networks (DSN)}, pages 391--402. IEEE, 2017.

\bibitem{G_201_sum}
Google.
\newblock Android security 2017 year in review, 2017.
\newblock
  \url{https://source.android.com/security/reports/Google\_Android\_Security\_2017\_Report_Final.pdf}.

\bibitem{lindorfer2014andradar}
Martina Lindorfer, Stamatis Volanis, Alessandro Sisto, Matthias
  Neugschwandtner, Elias Athanasopoulos, Federico Maggi, Christian Platzer,
  Stefano Zanero, and Sotiris Ioannidis.
\newblock Andradar: fast discovery of android applications in alternative
  markets.
\newblock In {\em International Conference on Detection of Intrusions and
  Malware, and Vulnerability Assessment}, pages 51--71. Springer, 2014.

\bibitem{pendlebury2018tesseract}
Feargus Pendlebury, Fabio Pierazzi, Roberto Jordaney, Johannes Kinder, and
  Lorenzo Cavallaro.
\newblock \$\{\$tesseract\$\}\$: Eliminating experimental bias in malware
  classification across space and time.
\newblock In {\em 28th USENIX Security Symposium}, pages 729--746, 2019.

\bibitem{goodfellow2014explaining}
Ian~J Goodfellow, Jonathon Shlens, and Christian Szegedy.
\newblock Explaining and harnessing adversarial examples.
\newblock {\em arXiv preprint arXiv:1412.6572}, 2014.

\bibitem{grosse2017adversarial}
Kathrin Grosse, Nicolas Papernot, Praveen Manoharan, Michael Backes, and
  Patrick McDaniel.
\newblock Adversarial examples for malware detection.
\newblock In {\em European Symposium on Research in Computer Security}, pages
  62--79. Springer, 2017.

\bibitem{kuppa2019black}
Aditya Kuppa, Slawomir Grzonkowski, Muhammad~Rizwan Asghar, and Nhien-An
  Le-Khac.
\newblock Black box attacks on deep anomaly detectors.
\newblock In {\em Proceedings of the 14th International Conference on
  Availability, Reliability and Security}, page~21. ACM, 2019.

\bibitem{yuan2019adversarial}
Xiaoyong Yuan, Pan He, Qile Zhu, and Xiaolin Li.
\newblock Adversarial examples: Attacks and defenses for deep learning.
\newblock {\em IEEE Transactions on Neural Networks and Learning Systems},
  2019.

\bibitem{10.1145/390013.808479}
Frances~E. Allen.
\newblock Control flow analysis.
\newblock {\em SIGPLAN Not.}, 5(7):1–19, jul 1970.

\bibitem{alzantot2018generating}
Moustafa Alzantot, Yash Sharma, Ahmed Elgohary, Bo-Jhang Ho, Mani Srivastava,
  and Kai-Wei Chang.
\newblock Generating natural language adversarial examples.
\newblock {\em arXiv preprint arXiv:1804.07998}, 2018.

\bibitem{apruzzese2018evading}
Giovanni Apruzzese and Michele Colajanni.
\newblock Evading botnet detectors based on flows and random forest with
  adversarial samples.
\newblock In {\em 2018 IEEE 17th International Symposium on Network Computing
  and Applications (NCA)}, pages 1--8. IEEE, 2018.

\bibitem{dang2017evading}
Hung Dang, Yue Huang, and Ee-Chien Chang.
\newblock Evading classifiers by morphing in the dark.
\newblock In {\em Proceedings of the 2017 ACM SIGSAC Conference on Computer and
  Communications Security}, pages 119--133, 2017.

\bibitem{demontis2017yes}
Ambra Demontis, Marco Melis, Battista Biggio, Davide Maiorca, Daniel Arp,
  Konrad Rieck, Igino Corona, Giorgio Giacinto, and Fabio Roli.
\newblock Yes, machine learning can be more secure! a case study on android
  malware detection.
\newblock {\em IEEE Transactions on Dependable and Secure Computing}, 2017.

\bibitem{li2018textbugger}
Jinfeng Li, Shouling Ji, Tianyu Du, Bo~Li, and Ting Wang.
\newblock Textbugger: Generating adversarial text against real-world
  applications.
\newblock {\em arXiv preprint arXiv:1812.05271}, 2018.

\bibitem{rastogi2013droidchameleon}
Vaibhav Rastogi, Yan Chen, and Xuxian Jiang.
\newblock Droidchameleon: evaluating android anti-malware against
  transformation attacks.
\newblock In {\em Proceedings of the 8th ACM SIGSAC Symposium on Information,
  Computer and Communications Security}, pages 329--334. ACM, 2013.

\bibitem{zheng2012adam}
Min Zheng, Patrick~PC Lee, and John~CS Lui.
\newblock Adam: an automatic and extensible platform to stress test android
  anti-virus systems.
\newblock In {\em International Conference on Detection of Intrusions and
  Malware, and Vulnerability Assessment}, pages 82--101. Springer, 2012.

\bibitem{pkdd2013}
Battista Biggio, Igino Corona, Davide Maiorca, Blaine Nelson, Nedim Srndic,
  Pavel Laskov, Giorgio Giacinto, and Fabio Roli.
\newblock Evasion attacks against machine learning at test time.
\newblock In {\em European Conference on Machine Learning and Knowledge
  Discovery in Databases}, pages 387--402, 2013.

\bibitem{carlini2017towards}
Nicholas Carlini and David Wagner.
\newblock Towards evaluating the robustness of neural networks.
\newblock In {\em 2017 IEEE Symposium on Security and Privacy}, pages 39--57.
  IEEE, 2017.

\bibitem{chen2018droideye}
Lingwei Chen, Shifu Hou, Yanfang Ye, and Shouhuai Xu.
\newblock Droideye: Fortifying security of learning-based classifier against
  adversarial android malware attacks.
\newblock In {\em 2018 IEEE/ACM International Conference on Advances in Social
  Networks Analysis and Mining}, pages 782--789. IEEE, 2018.

\bibitem{shahpasand2019adversarial}
Maryam Shahpasand, Len Hamey, Dinusha Vatsalan, and Minhui Xue.
\newblock Adversarial attacks on mobile malware detection.
\newblock In {\em 2019 IEEE 1st International Workshop on Artificial
  Intelligence for Mobile}, pages 17--20. IEEE, 2019.

\bibitem{shao2019multi}
Rui Shao, Xiangyuan Lan, Jiawei Li, and Pong~C Yuen.
\newblock Multi-adversarial discriminative deep domain generalization for face
  presentation attack detection.
\newblock In {\em Proceedings of the IEEE Conference on Computer Vision and
  Pattern Recognition}, pages 10023--10031, 2019.

\bibitem{xu2020towards}
Qiuling Xu, Guanhong Tao, Siyuan Cheng, Lin Tan, and Xiangyu Zhang.
\newblock Towards feature space adversarial attack.
\newblock {\em arXiv preprint arXiv:2004.12385}, 2020.

\bibitem{zikratov2017formalization}
Igor~A Zikratov, Victoria Korzhuk, Ilya Shilov, and Alexey Gvozdev.
\newblock Formalization of the feature space for detection of attacks on
  wireless sensor networks.
\newblock In {\em 2017 20th Conference of Open Innovations Association
  (FRUCT)}, pages 526--533. IEEE, 2017.

\bibitem{aydogan2015automatic}
Emre Aydogan and Sevil Sen.
\newblock Automatic generation of mobile malwares using genetic programming.
\newblock In {\em European Conference on the Applications of Evolutionary
  Computation}, pages 745--756. Springer, 2015.

\bibitem{chen2017adversarial}
Lingwei Chen, Shifu Hou, Yanfang Ye, and Lifei Chen.
\newblock An adversarial machine learning model against android malware evasion
  attacks.
\newblock In {\em Asia-Pacific Web and Web-Age Information Management Joint
  Conference on Web and Big Data}, pages 43--55. Springer, 2017.

\bibitem{hu2017generating}
Weiwei Hu and Ying Tan.
\newblock Generating adversarial malware examples for black-box attacks based
  on gan.
\newblock {\em arXiv preprint arXiv:1702.05983}, 2017.

\bibitem{ming2015replacement}
Jiang Ming, Zhi Xin, Pengwei Lan, Dinghao Wu, Peng Liu, and Bing Mao.
\newblock Replacement attacks: automatically impeding behavior-based malware
  specifications.
\newblock In {\em International Conference on Applied Cryptography and Network
  Security}, pages 497--517. Springer, 2015.

\bibitem{zhao2019unsupervised}
Guoping Zhao, Mingyu Zhang, Jiajun Liu, and Ji-Rong Wen.
\newblock Unsupervised adversarial attacks on deep feature-based retrieval with
  gan.
\newblock {\em arXiv preprint arXiv:1907.05793}, 2019.

\bibitem{chen2018android}
Xiao Chen, Chaoran Li, Derui Wang, Sheng Wen, Jun Zhang, Surya Nepal, Yang
  Xiang, and Kui Ren.
\newblock Android hiv: A study of repackaging malware for evading
  machine-learning detection.
\newblock {\em IEEE Transactions on Information Forensics and Security}, 2019.

\bibitem{pierazzi2020problemspace}
F.~Pierazzi, F.~Pendlebury, J.~Cortellazzi, and L.~Cavallaro.
\newblock Intriguing properties of adversarial ml attacks in the problem space.
\newblock In {\em 2020 IEEE Symposium on Security and Privacy}, pages
  1308--1325. IEEE Computer Society, 2020.

\bibitem{salem2018repackman}
Aleieldin Salem, F~Franziska Paulus, and Alexander Pretschner.
\newblock Repackman: A tool for automatic repackaging of android apps.
\newblock In {\em Proceedings of the 1st International Workshop on Advances in
  Mobile App Analysis}, pages 25--28. ACM, 2018.

\bibitem{berger2020evasion}
Harel Berger, Chen Hajaj, and Amit Dvir.
\newblock Evasion is not enough: A case study of android malware.
\newblock In {\em International Symposium on Cyber Security Cryptography and
  Machine Learning}, pages 167--174. Springer, 2020.

\bibitem{li2019android}
Chenglin Li, Keith Mills, Di~Niu, Rui Zhu, Hongwen Zhang, and Husam Kinawi.
\newblock Android malware detection based on factorization machine.
\newblock {\em IEEE Access}, 7:184008--184019, 2019.

\bibitem{AdversarialDeepEnsemble}
D.~{Li} and Q.~{Li}.
\newblock Adversarial deep ensemble: Evasion attacks and defenses for malware
  detection.
\newblock {\em IEEE Transactions on Information Forensics and Security},
  15:3886--3900, 2020.

\bibitem{droidapiminber_code}
ChenJunHero.
\newblock {droidapiminer code}.
\newblock github, 2018.
\newblock \url{https://github.com/ChenJunHero/DroidAPIMiner}.

\bibitem{maiorca2015stealth}
Davide Maiorca, Davide Ariu, Igino Corona, Marco Aresu, and Giorgio Giacinto.
\newblock Stealth attacks: An extended insight into the obfuscation effects on
  android malware.
\newblock {\em Computers \& Security}, 51:16--31, 2015.

\bibitem{meng2016mystique}
Guozhu Meng, Yinxing Xue, Chandramohan Mahinthan, Annamalai Narayanan, Yang
  Liu, Jie Zhang, and Tieming Chen.
\newblock Mystique: Evolving android malware for auditing anti-malware tools.
\newblock In {\em Proceedings of the 11th ACM on Asia Conference on Computer
  and Communications Security}, pages 365--376. ACM, 2016.

\bibitem{backes2016reliable}
Michael Backes, Sven Bugiel, and Erik Derr.
\newblock Reliable third-party library detection in android and its security
  applications.
\newblock In {\em Proceedings of the 2016 ACM SIGSAC Conference on Computer and
  Communications Security}, pages 356--367. ACM, 2016.

\bibitem{xu2018cdgdroid}
Zhiwu Xu, Kerong Ren, Shengchao Qin, and Florin Craciun.
\newblock Cdgdroid: Android malware detection based on deep learning using cfg
  and dfg.
\newblock In {\em International Conference on Formal Engineering Methods},
  pages 177--193. Springer, 2018.

\bibitem{zhiwu2019android}
XU~Zhiwu, Kerong Ren, and Fu~Song.
\newblock Android malware family classification and characterization using cfg
  and dfg.
\newblock In {\em 2019 International Symposium on Theoretical Aspects of
  Software Engineering (TASE)}, pages 49--56. IEEE, 2019.

\bibitem{ikram2019dadidroid}
Muhammad Ikram, Pierrick Beaume, and Mohamed~Ali Kaafar.
\newblock Dadidroid: An obfuscation resilient tool for detecting android
  malware via weighted directed call graph modelling.
\newblock {\em arXiv preprint arXiv:1905.09136}, 2019.

\bibitem{piao2016server}
Yuxue Piao, Jin-Hyuk Jung, and Jeong~Hyun Yi.
\newblock Server-based code obfuscation scheme for apk tamper detection.
\newblock {\em Security and Communication Networks}, 9(6):457--467, 2016.

\bibitem{sun2014nativeguard}
Mengtao Sun and Gang Tan.
\newblock Nativeguard: Protecting android applications from third-party native
  libraries.
\newblock In {\em Proceedings of the 2014 ACM Conference on Security and
  Privacy in Wireless \& Mobile Networks}, pages 165--176. ACM, 2014.

\bibitem{shabtai2010intrusion}
Asaf Shabtai, Uri Kanonov, and Yuval Elovici.
\newblock Intrusion detection for mobile devices using the knowledge-based,
  temporal abstraction method.
\newblock {\em Journal of Systems and Software}, 83(8):1524--1537, 2010.

\bibitem{saracino2016madam}
Andrea Saracino, Daniele Sgandurra, Gianluca Dini, and Fabio Martinelli.
\newblock Madam: Effective and efficient behavior-based android malware
  detection and prevention.
\newblock {\em IEEE Transactions on Dependable and Secure Computing},
  15(1):83--97, 2016.

\bibitem{wang2021android}
Xinning Wang and Chong Li.
\newblock Android malware detection through machine learning on kernel task
  structures.
\newblock {\em Neurocomputing}, 435:126--150, 2021.

\bibitem{kang2013android}
Byeongho Kang, BooJoong Kang, Jungtae Kim, and Eul~Gyu Im.
\newblock Android malware classification method: Dalvik bytecode frequency
  analysis.
\newblock In {\em Proceedings of the 2013 Research in Adaptive and Convergent
  Systems}, pages 349--350. 2013.

\bibitem{chen2018tinydroid}
Tieming Chen, Qingyu Mao, Yimin Yang, Mingqi Lv, and Jianming Zhu.
\newblock Tinydroid: a lightweight and efficient model for android malware
  detection and classification.
\newblock {\em Mobile Information Systems}, 2018, 2018.

\bibitem{damashek1995gauging}
Marc Damashek.
\newblock Gauging similarity with n-grams: Language-independent categorization
  of text.
\newblock {\em Science}, 267(5199):843--848, 1995.

\bibitem{yang2015appspear}
Wenbo Yang, Yuanyuan Zhang, Juanru Li, Junliang Shu, Bodong Li, Wenjun Hu, and
  Dawu Gu.
\newblock Appspear: Bytecode decrypting and dex reassembling for packed android
  malware.
\newblock In {\em International Symposium on Recent Advances in Intrusion
  Detection}, pages 359--381. Springer, 2015.

\bibitem{martin2019android}
Alejandro Mart{\'\i}n, Ra{\'u}l Lara-Cabrera, and David Camacho.
\newblock Android malware detection through hybrid features fusion and ensemble
  classifiers: The andropytool framework and the omnidroid dataset.
\newblock {\em Information Fusion}, 52:128--142, 2019.

\bibitem{wang2015reevaluating}
Haoyu Wang, Yao Guo, Zihao Tang, Guangdong Bai, and Xiangqun Chen.
\newblock Reevaluating android permission gaps with static and dynamic
  analysis.
\newblock In {\em 2015 IEEE Global Communications Conference (GLOBECOM)}, pages
  1--6. IEEE, 2015.

\bibitem{lindorfer2015marvin}
Martina Lindorfer, Matthias Neugschwandtner, and Christian Platzer.
\newblock Marvin: Efficient and comprehensive mobile app classification through
  static and dynamic analysis.
\newblock In {\em 2015 IEEE 39th annual computer software and applications
  conference}, volume~2, pages 422--433. IEEE, 2015.

\bibitem{ding2021hybrid}
Chao Ding, Nurbol Luktarhan, Bei Lu, and Wenhui Zhang.
\newblock A hybrid analysis-based approach to android malware family
  classification.
\newblock {\em Entropy}, 23(8):1009, 2021.

\bibitem{rosenberg2018generic}
Ishai Rosenberg, Asaf Shabtai, Lior Rokach, and Yuval Elovici.
\newblock Generic black-box end-to-end attack against state of the art api call
  based malware classifiers.
\newblock In {\em International Symposium on Research in Attacks, Intrusions,
  and Defenses}, pages 490--510. Springer, 2018.

\bibitem{cara2020feasibility}
Fabrizio Cara, Michele Scalas, Giorgio Giacinto, and Davide Maiorca.
\newblock On the feasibility of adversarial sample creation using the android
  system api.
\newblock {\em Information}, 11(9):433, 2020.

\bibitem{onwuzurike2019mamadroidold}
Lucky Onwuzurike, Enrico Mariconti, Panagiotis Andriotis, Emiliano~De
  Cristofaro, Gordon Ross, and Gianluca Stringhini.
\newblock Mamadroid: Detecting android malware by building markov chains of
  behavioral models (extended version).
\newblock {\em ACM Transactions on Privacy and Security}, 22(2):14, 2019.

\bibitem{brooks2011handbook}
Steve Brooks, Andrew Gelman, Galin Jones, and Xiao-Li Meng.
\newblock {\em Handbook of markov chain monte carlo}.
\newblock CRC press, 2011.

\bibitem{geyer1992practical}
Charles~J Geyer.
\newblock Practical markov chain monte carlo.
\newblock {\em Statistical science}, pages 473--483, 1992.

\bibitem{marjoram2003markov}
Paul Marjoram, John Molitor, Vincent Plagnol, and Simon Tavar{\'e}.
\newblock Markov chain monte carlo without likelihoods.
\newblock {\em Proceedings of the National Academy of Sciences},
  100(26):15324--15328, 2003.

\bibitem{mama_implementation}
Enrico Mariconti, Lucky Onwuzurike, Panagiotis Andriotis, Emiliano
  De~Cristofaro, Gordon Ross, and Gianluca Stringhini.
\newblock {MaMaDroid implementation}.
\newblock bitbucket, 2018.
\newblock
  \url{https://bitbucket.org/gianluca_students/mamadroid_code/src/master/}.

\bibitem{spooren2019use}
Jan Spooren, Davy Preuveneers, Lieven Desmet, Peter Janssen, and Wouter Joosen.
\newblock On the use of dgas in malware: an everlasting competition of
  detection and evasion.
\newblock {\em ACM SIGAPP Applied Computing Review}, 19(2):31--43, 2019.

\bibitem{allix2016androzoo}
Kevin Allix, Tegawend{\'e}~F Bissyand{\'e}, Jacques Klein, and Yves Le~Traon.
\newblock Androzoo: Collecting millions of android apps for the research
  community.
\newblock In {\em 2016 IEEE/ACM 13th Working Conference on Mining Software
  Repositories}, pages 468--471. IEEE, 2016.

\bibitem{ali2016aspectdroid}
Aisha Ali-Gombe, Irfan Ahmed, Golden~G Richard~III, and Vassil Roussev.
\newblock Aspectdroid: Android app analysis system.
\newblock In {\em Proceedings of the Sixth ACM Conference on Data and
  Application Security and Privacy}, pages 145--147, 2016.

\bibitem{frenklach2021android}
Tatiana Frenklach, Dvir Cohen, Asaf Shabtai, and Rami Puzis.
\newblock Android malware detection via an app similarity graph.
\newblock {\em Computers \& Security}, 109:102386, 2021.

\bibitem{maiorca2017r}
Davide Maiorca, Francesco Mercaldo, Giorgio Giacinto, Corrado~Aaron Visaggio,
  and Fabio Martinelli.
\newblock R-packdroid: Api package-based characterization and detection of
  mobile ransomware.
\newblock In {\em Proceedings of the symposium on applied computing}, pages
  1718--1723, 2017.

\bibitem{yuan2020byte}
Baoguo Yuan, Junfeng Wang, Dong Liu, Wen Guo, Peng Wu, and Xuhua Bao.
\newblock Byte-level malware classification based on markov images and deep
  learning.
\newblock {\em Computers \& Security}, 92:101740, 2020.

\bibitem{zulkifli2018android}
Aqil Zulkifli, Isredza Rahmi~A Hamid, Wahidah~Md Shah, and Zubaile Abdullah.
\newblock Android malware detection based on network traffic using decision
  tree algorithm.
\newblock In {\em International Conference on Soft Computing and Data Mining},
  pages 485--494. Springer, 2018.

\bibitem{tong2017framework}
Liang Tong, Bo~Li, Chen Hajaj, Chaowei Xiao, Ning Zhang, and Yevgeniy
  Vorobeychik.
\newblock Improving robustness of \$\{\$ml\$\}\$ classifiers against realizable
  evasion attacks using conserved features.
\newblock In {\em 28th USENIX Security Symposium}, pages 285--302, 2019.

\bibitem{brama2022evaluation}
Haya Brama, Lihi Dery, and Tal Grinshpoun.
\newblock Evaluation of neural networks defenses and attacks using ndcg and
  reciprocal rank metrics, 2022.

\bibitem{nguyen2020ensemble}
Tien~Thanh Nguyen, Anh~Vu Luong, Manh~Truong Dang, Alan Wee-Chung Liew, and
  John McCall.
\newblock Ensemble selection based on classifier prediction confidence.
\newblock {\em Pattern Recognition}, 100:107104, 2020.

\bibitem{arora2019permpair}
Anshul Arora, Sateesh~K Peddoju, and Mauro Conti.
\newblock Permpair: Android malware detection using permission pairs.
\newblock {\em IEEE Transactions on Information Forensics and Security}, 2019.

\bibitem{aswini2014droid}
AM~Aswini and P~Vinod.
\newblock Droid permission miner: Mining prominent permissions for android
  malware analysis.
\newblock In {\em The Fifth International Conference on the Applications of
  Digital Information and Web Technologies}, pages 81--86. IEEE, 2014.

\bibitem{li2018significant}
Jin Li, Lichao Sun, Qiben Yan, Zhiqiang Li, Witawas Srisa-an, and Heng Ye.
\newblock Significant permission identification for machine-learning-based
  android malware detection.
\newblock {\em IEEE Transactions on Industrial Informatics}, 14(7):3216--3225,
  2018.

\bibitem{sanz2013puma}
Borja Sanz, Igor Santos, Carlos Laorden, Xabier Ugarte-Pedrero, Pablo~Garcia
  Bringas, and Gonzalo {\'A}lvarez.
\newblock Puma: Permission usage to detect malware in android.
\newblock In {\em International Joint Conference CISIS'12-ICEUTE 12-SOCO 12
  Special Sessions}, pages 289--298. Springer, 2013.

\bibitem{Android_permissions_cust}
Google.
\newblock Define a custom app permission.
\newblock Android Developers website, 2022.
\newblock
  \url{https://developer.android.com/guide/topics/permissions/defining}.

\end{thebibliography}
